\documentclass[11pt]{article}
\usepackage{amsmath,amssymb,color,graphics,epsfig,cite}
\usepackage{ulem}
\usepackage{amsfonts}
\newcommand{\hoch}[1]{$\, ^{#1}$}

\makeatletter
\@addtoreset{equation}{section}
\makeatother

\newcommand{\be}{\begin{equation}}
\newcommand{\ee}{\end{equation}}
\newcommand{\bea}{\setlength\arraycolsep{2pt} \begin{eqnarray}}
\newcommand{\eea}{\end{eqnarray}}
\newcommand{\nn}{\nonumber}

\def\ft#1#2{{\textstyle{\frac{\scriptstyle #1}{\scriptstyle #2} } }}
\def\fft#1#2{{\frac{#1}{#2}}}

\def\0{{\sst{(0)}}}
\def\1{{\sst{(1)}}}
\def\2{{\sst{(2)}}}
\def\3{{\sst{(3)}}}
\def\4{{\sst{(4)}}}
\def\5{{\sst{(5)}}}
\def\6{{\sst{(6)}}}
\def\7{{\sst{(7)}}}
\def\8{{\sst{(8)}}}
\def\sst#1{{\scriptscriptstyle #1}}

\begin{document}

	%\begin{flushright}
	%\hfill{MI-TH-1533}
	%\end{flushright}
	
	%\vspace{25pt}
	\begin{center}
		{\Large {\bf Charged Ellis Wormhole and Black Bounce}}

		\vspace{40pt}
	
		{\large Hyat Huang\hoch{1},
			Jinbo Yang\hoch{2*}}
		
		\vspace{15pt}
		
		\hoch{1}{\it Department of Physics, Beijing Normal University, Beijing 100875, China}
	
		\vspace{10pt}
		
    	\hoch{2}{\it Institute for Theoretical Physics, Kanazawa University, Kanazawa 920-1192, Japan}

		\vspace{40pt}
		
		\underline{ABSTRACT}
	\end{center}

By replacing the scalar $\phi$ with $i\phi$ in the solution constructed in Ref\cite{Huang:2019lsl}, we obtain electrically-charged wormhole and black hole solutions in the Einstein-Maxwell-scalar theory, in which the scalar is a phantom field coupled to the Maxwell field non-minimally. Our solutions are simpler than the previously-known charged wormholes in the Einstein-Maxwell-dilaton theory in that both the scalar and the radius of the foliating sphere of our solutions are independent of electric charge. The wormhole solution has two asymptotically flat regions and reduces to Ellis wormhole when the charge is zero. We draw the embedding diagrams for the wormholes and demonstrate that the two sides are asymmetric. As the charge increases, horizons will appear, and the wormhole becomes a black hole, with no curvature singularity but a wormhole throat or a bounce. We analyze black hole thermodynamics and the causal structure. We also determine the photon spheres of both the wormhole and the black hole and discuss their observable characteristics.

\vfill {\footnotesize Hyat Huang: hyat@mail.bnu.edu.cn \qquad Jinbo Yang: j\_yang@hep.stu.kanazawa.ac.jp }

	\thispagestyle{empty}
	
	\pagebreak

	\tableofcontents
	\addtocontents{toc}{\protect\setcounter{tocdepth}{2}}
	
	%%%%%%%%%%%%%%%%%%%%%%%%%%%%%%%%%%%%%%%%

	\newpage
	%%%%%%%%%%%%%%%%%%%%%%%%%%%%%%%%%%%%%%%%
\section{Introduction}

Wormhole physics is an important topic in gravity and cosmology, since the wormhole acts as a tunnel connecting two far separated regions of the same universe or even different universes. The paper written by Ludwig Flamm \cite{Flamm} in 1916 gave the first proposal of a wormhole. The first serious work on wormholes, the proposition of Einstein-Rosen bridge, was conducted by Einstein and Rosen in 1935 \cite{ER_bridge}. However, the Einstein-Rosen bridge is not traversable, even for a photon. Morris and Thorne investigated several properties of traversable wormholes \cite{Morris:1988cz} and Ellis constructed the famous eponymous wormhole solution in Einstein gravity coupled to a free phantom scalar \cite{Ellis:1973yv}.

The phantom scalar involved in the Ellis wormhole has always been the cause of the controversy in the subject. A phantom field is defined by flapping the sign of its kinetic term in the Lagrangian, which usually  leads to instabilities of the system. In fact, Morris and Thorne \cite{Morris:1988cz} showed that, in the framework of Einstein gravity, maintaining wormhole throat needs exotic matter that violates the null energy condition (NEC). Thus the existence of a traversable wormhole in nature or the possibility of constructing have been put in doubt because there has been no evidence of matter that violates the NEC. However, the NEC could be violated in early universe, and it has been an ongoing research topic on early cosmology. Papers have shown that wormholes could originate from the quantum fluctuations in the Plankian era of the early universe\cite{Roman:1992xj}, and then grew via inflation to the detectable scale that might be detectable \cite{Jusufi:2017vta}.

Meanwhile, it was argued that the instability problem of a phantom field could be cured \cite{Piazza:2004df}. Furthermore, the cosmological observations suggest that dark energy contributes around 70\% of the total energy in our Universe. The possibility of dark energy is a phantom field has also been also raised \cite{Caldwell:1999ew,Nakonieczna:2015apa,Nakonieczna:2012in}. It is worth mentioning that phantom fields can also arise in string theory, as the negative tension branes, which play an important role in string dualities \cite{Hull:1998ym,Vafa:2001qf,Okuda:2006fb}. These imply that phantom fields arise from diverse subjects in physics and may not be simply ignored. In fact quantum field theory allows the violations of NEC \cite{Tipler:1978zz,Borde:1987qr}, and a famous example is the Casimir effect. The phantom field that arises from quantum effects may also serve as a tool to study quantum gravity.

The construction of the analytic wormhole solutions in various theories become quite prominent nowadays \cite{Goulart:2017iko,Maldacena:2018gjk,Carvente:2019gkd,Baruah:2019cfg,Sahoo:2018kct,Chew:2019lsa,Miranda:2019uer,Garattini:2019ivd,Garattini:2019wka,Myrzakulov:2015kda}. In particular, the paper\cite{Goulart:2017iko} generalizes the Ellis wormhole to carry electric charges in the Einstein-Maxwell dilaton (EMD) theory where the dilaton is phantomlike. The wormhole solutions have attracted considerable attentions since they provide an analytic tool to study interstellar travel, time travel, the inspection of a black hole interior and the gravitational lensing of horizonless compact objects.

Black hole solutions can also arise from the EMD theories with the phantom dilaton. Gibbons and Rasheed \cite{Gibbons:1996pd} showed that there were massless black hole solutions in such theories. Since then many phantom black hole solutions have been constructed in a variety of theories \cite{Bronnikov:2005gm,Gao:2006iw,Jamil:2008bc,Clement:2009ai,AzregAinou:2011rj,Zhang:2017tbf}. These black holes have some interesting properties that are significantly different from the usual black holes that do satisfy the proper energy condition. For an example, the spacetime singularity inside the horizon can be avoided, replaced by a wormhole throat or more precisely a bounce. Regular black hole without singularity was proposed by Bardeen \cite{Bardeen:1968} in 1968 where the dominant energy condition was violated.  Inspired by the idea of regular black hole, Alex Simpson, Prado Martin-Moruno and Matt Visser\cite{Simpson:2018tsi,Simpson:2019cer} proposed the mechanism of `black bounce' recently. They combined the metrics of the Schwarzschild black hole and the Morris-Thorne traversable wormhole, and proposed a new NEC-violating metric
\be\label{viss}
ds^2=-(1-\ft{2m}{\sqrt{r^2+a^2}})dt^2+\ft{dr^2}{1-\ft{2m}{\sqrt{r^2+a^2}}}+(r^2+a^2)d\Omega_2^2,
\ee
where
\be
d\Omega_{2}^2=d\theta^2 + \sin^2\theta d\psi^2\,,
\ee
is the metric for the unit round 2-sphere.  The metric \eqref{viss} was shown to realize a bounce from a universe to another universe or a wormhole. A similar concept ``black universe'' is proposed by Bronnikov in Ref \cite{Bronnikov:2006fu}.

It is well-known that gravitational lensing is an interesting phenomenon whereby the propagation of light is affected by gravitating mass such as some massive compact objects. It is a useful tool to detect the mass distributions and test gravity theories.  As for black holes and wormholes, an intriguing aspect about gravitational lensing is that it gives rise to detectable shadows, which are casted by the `photon sphere' \cite{Claudel:2000yi,Izumi:2013tya,Tsukamoto:2016jzh,Shaikh:2018oul,Bronnikov:2018nub,Shaikh:2019jfr}.
Photon spheres are typically unstable, but black holes satisfying the dominant energy condition can also allow to have stable photon sphere \cite{Liu:2019rib}.

In this paper, we shall construct new classes of analytic wormhole and black hole solutions in more general Einstein-Maxwell-scalar (EMS) theories, keeping in mind the applications discussed above. The scalar is phantomlike, but the electromagnetic field in the theory may not be phantomlike at the asymptotic regions. When the charge is sufficiently small, the solution describes a wormhole.  The solution develops two horizons for sufficiently large charge, but there is no spacetime singularity. There are two types of black holes arising depending on the parameters. One is similar to the RN black hole except with the singularity is replaced by a wormhole throat. In other words, the wormhole throat is outside both of the two horizons. In the other case, the wormhole throat or more precisely the ``bounce'' occurs in between the two horizons and the solution can be viewed as a ``black hole-white hole pair''. This provides a concrete example of the black bounce in between the two horizons.

In order to study the observable characteristics, we compute and analyze the photon sphere of our solutions. We find that there can be either one or two photon spheres for the wormhole solutions. For the black hole solutions, there are always two photon spheres, with one outside of each horizon.

The paper is organized as follows. In section \ref{sec1}, we review the theory and the solutions constructed in Ref\cite{Huang:2019lsl} and present new solutions by making the transformation of the scalar $\phi$ to $i\phi$, together with some appropriate changes of the parameters in the theory. We then make a coordinate change so that the solution can describe the whole spacetime. In section \ref{sec3}, we study the wormhole solutions. The embedding diagrams of the wormholes with different parameters are drawn, which illustrate that the two parts of the wormhole are asymmetric. In section \ref{sec4}, we show that black holes emerge if we increase the value of charge $Q$. These black holes have no spacetime singularity and can be regarded as a specific example of black bounce. We also study the black hole thermodynamics. In section \ref{cs}, we investigate the causal structures of the solutions and draw their Carter-Penrose diagrams. We investigate the photon spheres of both the wormholes and the black holes in section \ref{sec5}. Finally the paper is concluded in section \ref{sec6}.

\section{The theory and solution}\label{sec1}

The Lagrangian for the theory in Ref\cite{Huang:2019lsl} is given by
\bea\label{lag0}
&&\mathcal{L}=\sqrt{-g}(R-\ft{1}{2}(\partial \phi)^2-\ft{1}{4}Z^{-1}F^2),\nn\\
&&Z=\gamma_1 \cosh \phi-\gamma_2 \sinh \phi,
\eea
where $Z$ is the coupling function of the scalar field and the Maxwell field $F=dA$. Under the transformation $(\phi, \gamma_2)$ to $(i\phi,-i\gamma_2)$, the Lagrangian remains real and becomes
\bea\label{lag}
&&\mathcal{L}=\sqrt{-g}(R+\ft{1}{2}(\partial \phi)^2-\ft{1}{4}Z^{-1}F^2).\nn\\
&&Z=\gamma_1 \cos \phi+\gamma_2 \sin\phi.
\eea
In this new theory, the scalar is a phantom field, and furthermore since $Z$ becomes a harmonic function in $\phi$, the Maxwell field can also become phantomlike since the sign of the coupling function $Z$ is not positive definite regardless the values of $\gamma_1$ and $\gamma_2$.

For a spherically symmetric metric, we present the solution here\footnote{The theory and the solution can also be derived by a step-by-step generalization of the Ellis wormhole theory, presented in the Appendix A in Ref\cite{Huang:2019lsl}.}
\bea\label{solution}
&&ds^2=-h dt^2+(\sigma^2 h)^{-1}d\rho^2+\rho^2d\Omega_{2}^2\quad  \quad A=\xi(\rho,q,Q)dt,\nn\\
&&h=1-\ft{\gamma_2 Q^2\sigma}{4q\rho}+\ft{\gamma_1 Q^2}{4 \rho^2},\qquad \sigma=\sqrt{1-\ft{q^2}{\rho^2}},\nn\\
&&\phi=2\arcsin(\ft{q}{\rho}),\qquad \xi'=\ft{QZ}{\sigma \rho^2}.
\eea
It might appear to have a divergence in the Lagrangian \eqref{lag}, when $Z$ vanishes; however,
the solution is not divergent. The solution from the Maxwell equation implies that
\be
F^2=-\ft{2Z^2Q^2}{\rho^4},
\ee
where $Z^2$ appears in the numerator and hence there is no divergence at $Z=0$.

In fact, the metric \eqref{solution} covers only a part of the whole spacetime since $\rho$ takes the range $\rho\in[q,\infty)$. To describe the whole spacetime, we introduce a new radial coordinate $r$, defined by
$r^2=\rho^2-q^2$. The solution now becomes
\bea\label{ax}
&&ds^2=-h dt^2+(h)^{-1}dr^2+(r^2+q^2)d\Omega_{2}^2\quad \phi=\phi(r,q), \quad A=\xi(r,q,Q)dt,\nn\\
&&h=1-\ft{\gamma_2 Q^2 r}{4q(r^2+q^2)}+\ft{\gamma_1 Q^2}{4 (r^2+q^2)},\qquad \phi=2\arccos(\ft{r}{\sqrt{r^2+q^2}}),\qquad \xi'=\ft{QZ}{r^2+q^2},\nn\\
&&Z=\frac{-\gamma_1 q^2+2 \gamma_2 q r+\gamma_1 r^2}{q^2+r^2},
\eea
where $r$ runs from $-\infty$ to $+\infty$, corresponding to asymptotic flat regions.
It's easy to check the solution \eqref{ax} satisfies all the E.O.Ms \eqref{eom}. The coupling function $Z$ is a rational polynomial in terms of $r$. It is worth noting that
at the wormhole throat $r=0$, we have $Z=-\gamma_1$.  At the asymptotic regions, we have instead
\be\label{inZ}
Z(\pm \infty)=\gamma_1+\ft{2q\gamma_2}{r}+...\,.
\ee
Thus the Maxwell field will become phantomlike either in the wormhole throat or the asymptotic regions.  In most of our solutions, we choose $\gamma_1>0$ and consequently the Maxwell field is normal in the asymptotic region.

The solution (\ref{ax}) is asymptotically flat so the mass of the solution can be obtained by expanding $h(r)$ in asymptotic regions, which are given by
\begin{equation}\label{mass}
M_{+}=\frac{\gamma_2 Q^2}{8q}  \;,\qquad\; M_{-}=-\frac{\gamma_2 Q^2}{8q}.
\end{equation}
In which $M_{+}$ is the mass viewed from the $r>0$ side, while $M_{-}$  is the mass for the $r<0$ side. Obviously, they are equal but opposite, satisfying $M_{+}+ M_{-}=0$, which signals that the two spacetime regions connected by the wormhole are asymmetric. The charge of the solution, which is related to the parameter $Q$, can be obtained from the standard flux integration
\begin{equation}\label{charge}
Q_{e}=\lim_{r\rightarrow\pm\infty }\frac{1}{4\pi}\int_{\Omega} F_{rt}\sqrt{g_{\theta\theta}g_{\phi\phi}}d\theta d\phi
=\lim_{r\rightarrow\pm\infty }Z(r)Q=\gamma_1Q.
\end{equation}
It is worth noting that the electric charge $Q_e$ is the same viewed in both $r>0$ and $r<0$ regions.
It enters the solution only through the metric function $h$. The scalar and the radius of the foliating sphere are independent of $Q_e$, and this is arguably the simplest way to charge the Ellis wormhole, compared to the similar class of solutions in literature\cite{Goulart:2017iko}.

\section{Wormhole solutions}\label{sec3}

\subsection{Charged Ellis wormhole}

The solution \eqref{ax} characterizes a static charged wormhole for some appropriate parameters. To guarantee the traversability of wormholes we need to ensure that $-g_{tt}=h$ is positive everywhere. It requires that
\be\label{conwh}
Q^2<\ft{8q^2(\gamma_1+\sqrt{\gamma_1^2+\gamma_2^2})}{\gamma_2^2},
\ee
or in terms of $(Q_e, M)$,
\be
 (\ft{Q_e}{\gamma_1})^2>\ft{8M^2(-\gamma_1+\sqrt{\gamma_1^2+\gamma_2^2})}{\gamma_2^2}.
\ee
In other words, mass should be sufficiently small compared to the electric charge.

\subsection{Properties of the wormhole solution}

The wormhole throat is located at $r=0$ and it connects the two asymptotically flat regions. The wormhole mass and charge are given by \eqref{mass} and \eqref{charge} respectively. The area of the wormhole throat is given by
\be
A=4\pi q^2.
\ee
It shows that the area of the throat is the same as the Ellis wormhole solution.  It depends only on $q$, the coefficient of the long-range $1/r$ falloff of the massless scalar $\phi$.  It is independent of the new parameter $Q$. The curvature of those regions, near to the wormhole throat, is influenced by the charge $Q$. It is easy to see that the two sides of the wormhole are different in general. The charged wormhole reduces to the Ellis wormhole when we turn off $Q$, i.e.
\bea\label{ellis1}
&&ds^2=-dt^2+dr^2+(r^2+q^2)d\Omega_{2}^2,\nn\\
&&\phi=2\arccos(\ft{r}{\sqrt{r^2+q^2}}).
\eea
\begin{figure}[h]
\centering	
 \includegraphics[width=9cm]{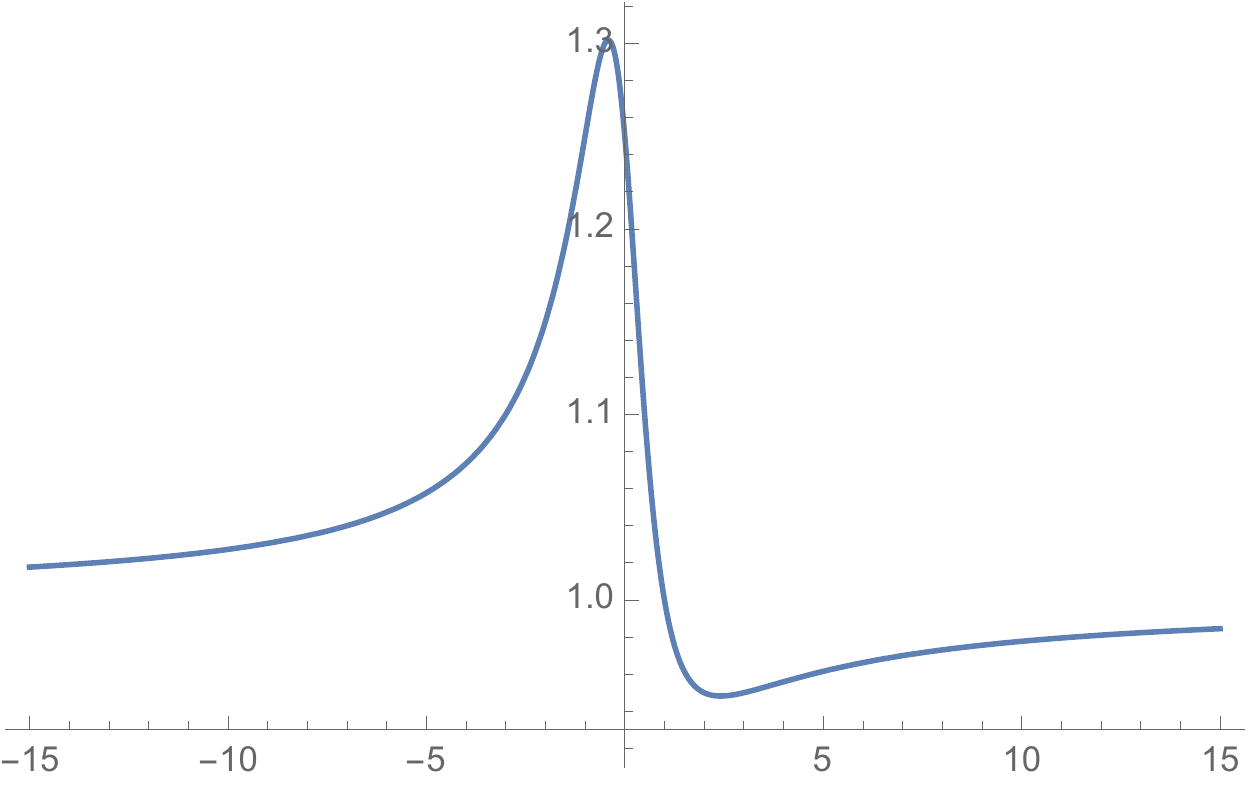}

	\caption{We show $h=h(r)$  with $q=1,\gamma_1=1,\gamma_2=1,Q=1$. The parameters satisfy \eqref{conwh} and indeed it shows that $h$ is positive and finite everywhere.}
	\label{fig:h}
\end{figure}

We compute the Ricci scalar and Kretschmann scalar $K:=R_{\mu\nu\rho\sigma}R^{\mu\nu\rho\sigma}$:
\bea\label{cur}
R&=&-\ft{q(4q^3+4qr^2+q \gamma_1 Q^2-\gamma_2 Q^2 r)}{2(q^2+r^2)^3},\nn\\
K&=&\ft{1}{4q^2(q^2+r^2)^6}\big(48 q^{10}+16q^8(6r^2+\gamma_1 Q^2)+q^6 (48r^4-8Q^2\gamma_1 r^2+3 Q^4\gamma_1^2)-24q^7 Q^2\gamma_2 r\nn\\
&&-12 q Q^4 \gamma_1\gamma_2 r^5-4q^5Q^2\gamma_2(2r^2+3\gamma_1Q^2)r+2\gamma_2 q^3 Q^2 r^3(8r^2+15Q^2\gamma_1)+3Q^4\gamma_2^2 r^6\nn\\
&&+2q^2 Q^4 r^4(7\gamma_1^2-5\gamma_2^2)-2q^4 Q^2 r^2(12r^2\gamma_1+Q^2(5\gamma_1^2-7\gamma_2^2))\big).
\eea
It is straightforward to check that the spacetime has no singularity if $q\neq0$. We would like to present the values of these two quantities at the wormhole throat $r=0$, given by
\bea
&&R(0)=-\ft{4q^3+q\gamma_1 Q^2}{2q^5},\nn\\
&&K(0)=\ft{12}{q^4}+\ft{4\gamma_1 Q^2}{q^6}+\ft{3\gamma_1^2 Q^4}{4q^8}.
\eea
Thus we obtain charged wormholes in EMS theory. As we shall discuss in section \ref{sec4}, if we increase the parameter $Q$ while keeping the wormhole throat $q$ fixed, event horizons can develop and the solutions can describe black holes without curvature singularities.

\subsection{Embedding diagram and flare-out condition}

\begin{figure}[h]
	  \includegraphics[width=5cm]{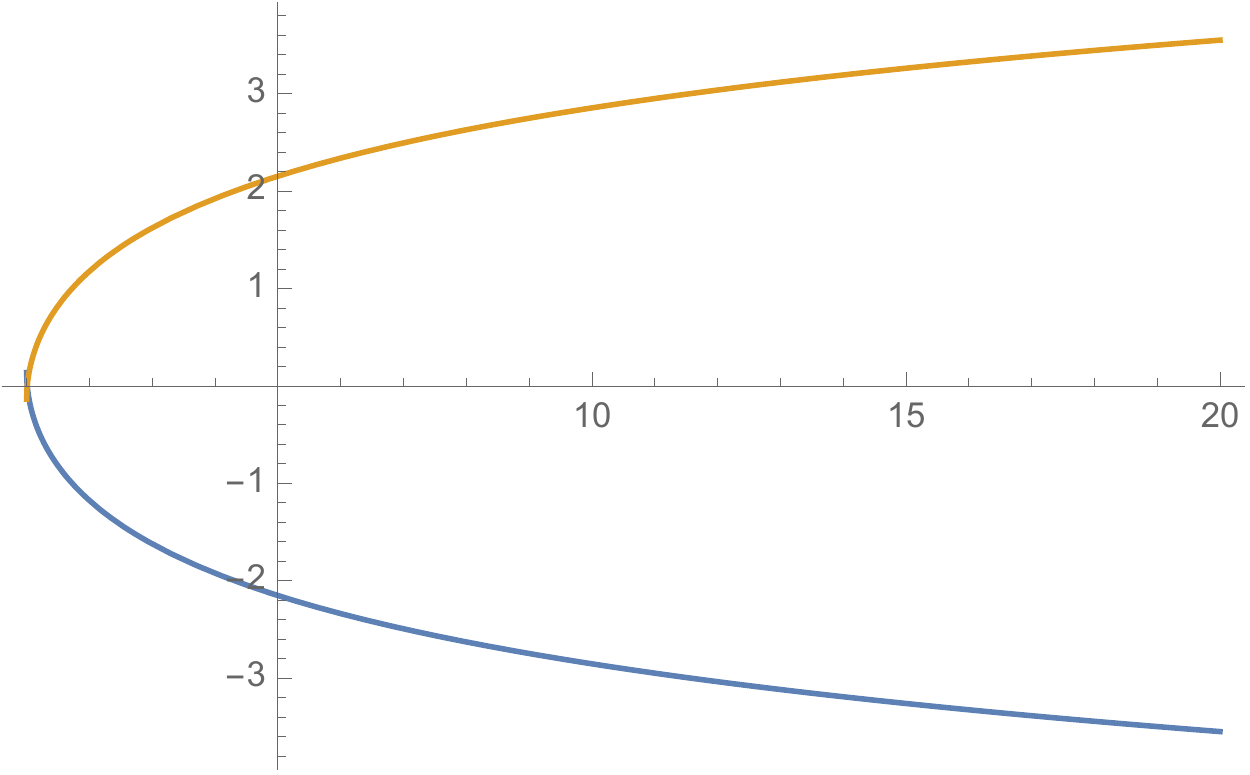} \ \ \
	\includegraphics[width=11cm]{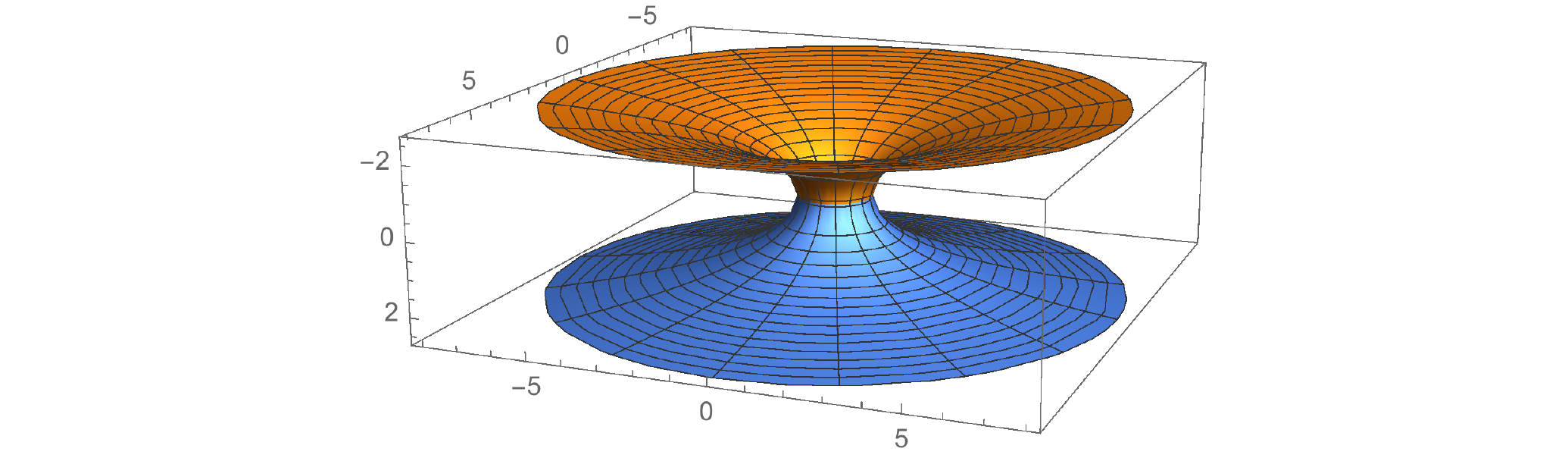} \\
	\ \ \ \ \ \ \ \includegraphics[width=5cm]{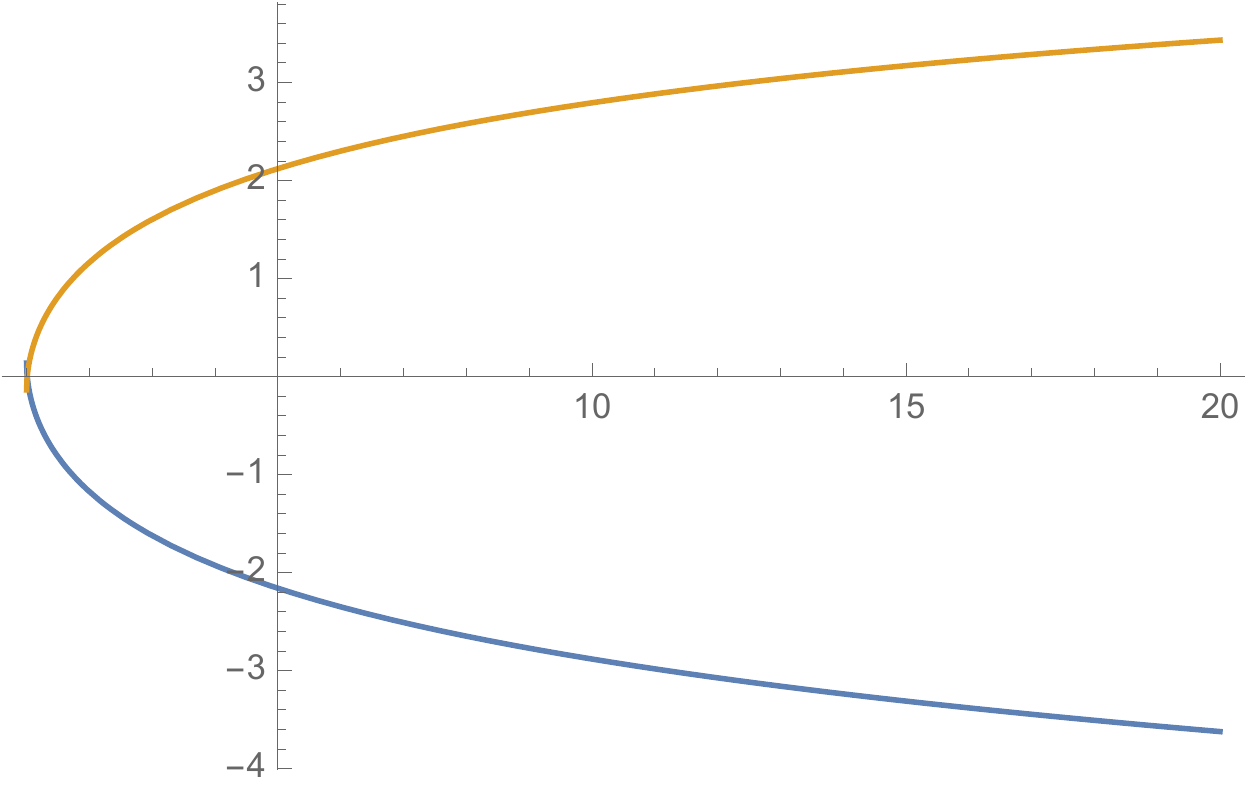} \ \ \
	\includegraphics[width=11cm]{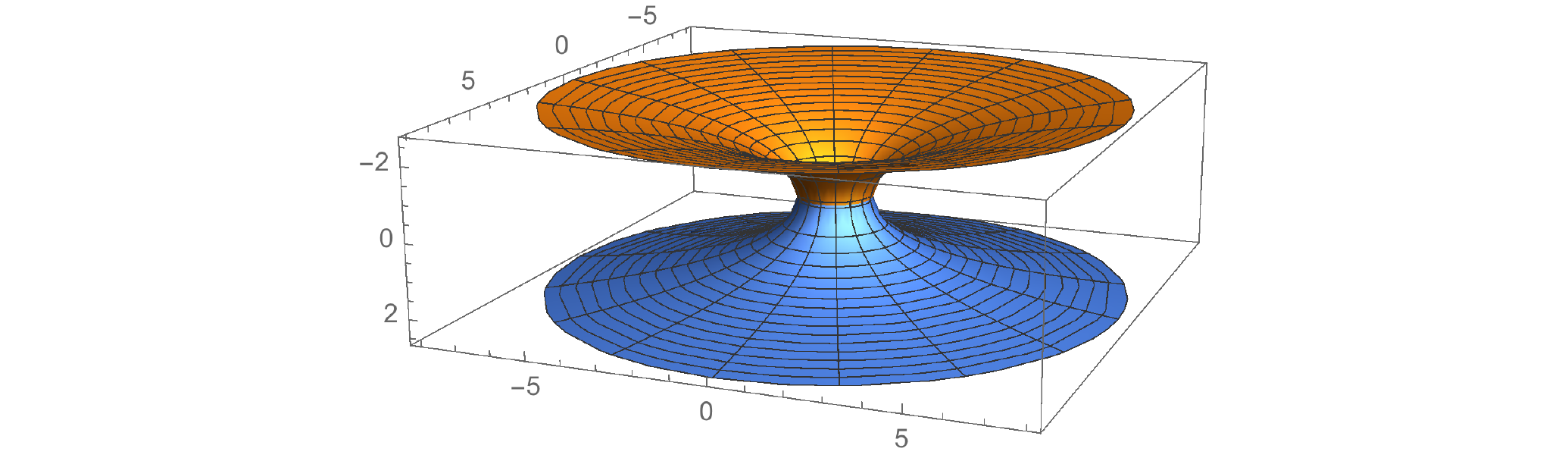} \\
		\ \ \ \ \ \includegraphics[width=5cm]{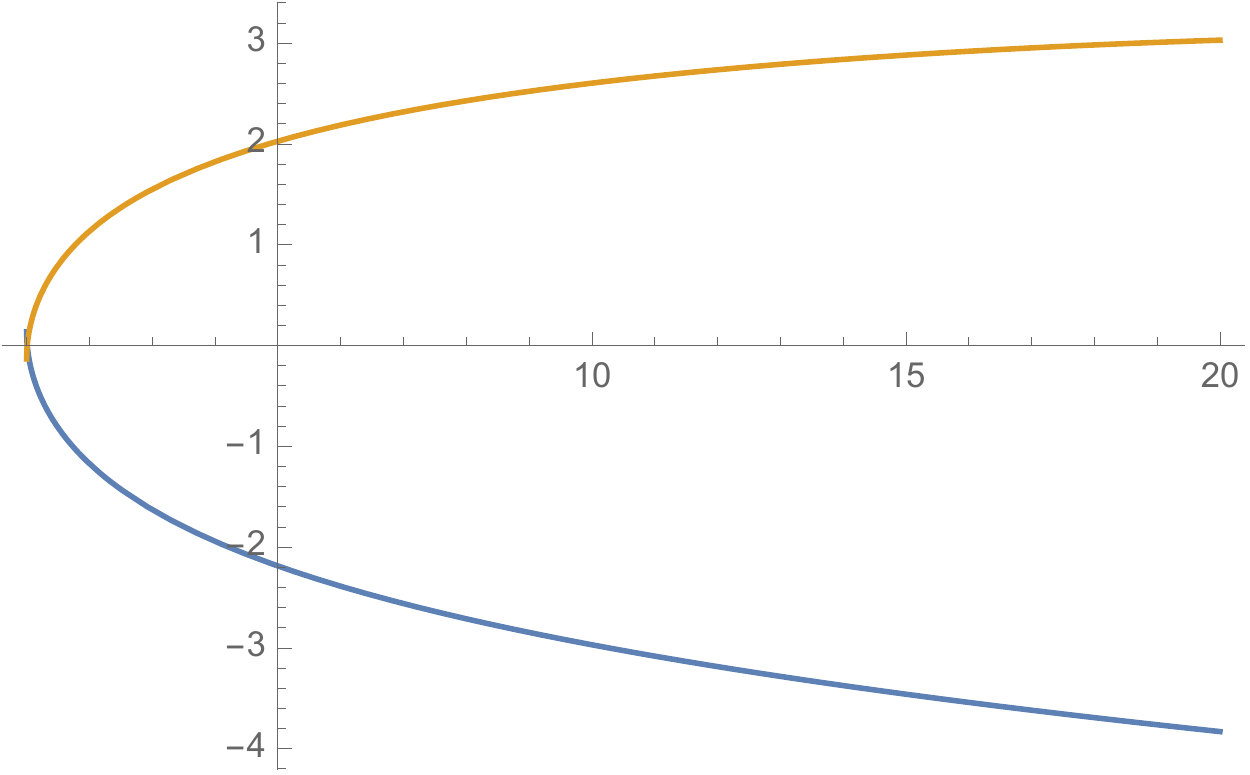} \ \ \
	\includegraphics[width=11cm]{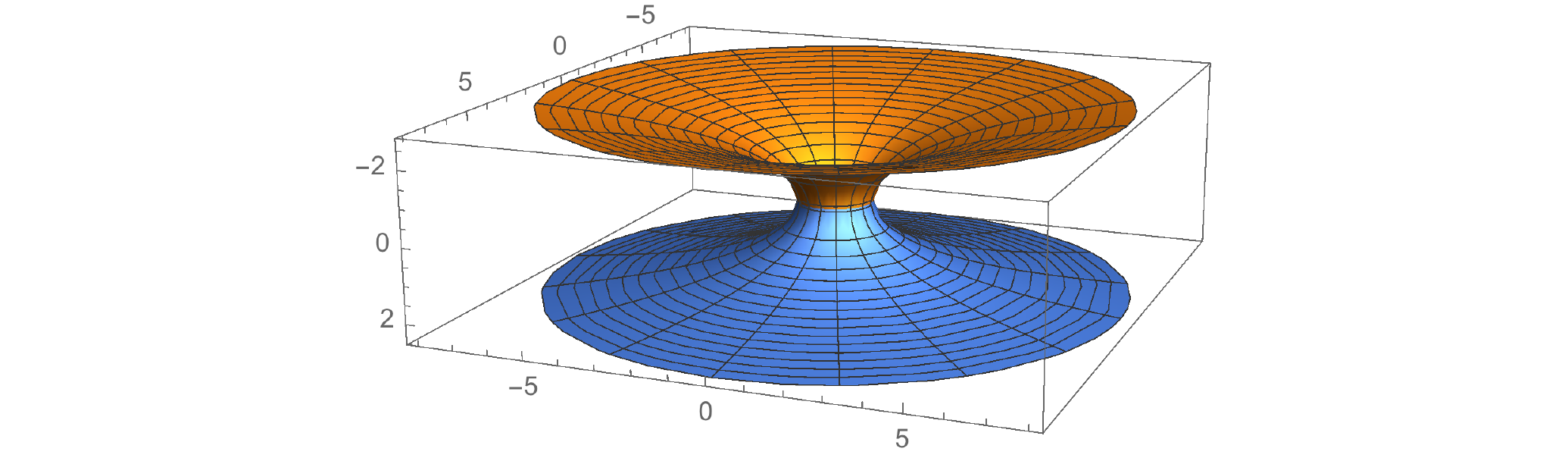}
	\caption{We fix $q=1,\gamma_1=1,\gamma_2=1$ and show $z=z(\rho)$ and the corresponding embedding diagrams with $Q=0,0.2,0.4$ respectively. The throats of wormhole are both located at $\rho=1$. The two regions of the spacetime are symmetric when $Q=0$. In fact, this case reduces to Ellis wormhole. The two regions of the spacetime are not symmetric when $Q\neq0$}
	\label{fig:wh2}
\end{figure}

Since the charged wormhole is asymmetric, it is necessary to split the metric (2.5) for constructing its embedding diagram. By using the coordinate transformation $r=\sqrt{\rho^2-q^2}$, one side of wormhole throat, namely $r>0$ in the \eqref{ax}, can be depicted by
\bea\label{solution2}
&&ds^2=-h_+ dt^2+(\sigma^2 h_+)^{-1}d\rho^2+\rho^2d\Omega_{2}^2,\nn\\
&&h_+=1-\ft{\gamma_2 Q^2\sigma}{4q\rho}+\ft{\gamma_1 Q^2}{4 \rho^2},\qquad \sigma=\sqrt{1-\ft{q^2}{\rho^2}},\qquad \rho\ge q.
\eea
In fact, \eqref{solution2} is the same as the original solution \eqref{solution}. On the other hand, by using the coordinate transformation $r=-\sqrt{\rho^2-q^2}$, we obtain the solution of the other side of the wormhole throat, namely $r<0$ in the \eqref{ax}. It is given by
\bea\label{theo2}
&&ds^2=-h_- dt^2+(\sigma^2 h_-)^{-1}d\rho^2+\rho^2d\Omega_{2}^2,\nn\\
&&h_-=1+\ft{\gamma_2 Q^2\sigma}{4q\rho}+\ft{\gamma_1 Q^2}{4 \rho^2}, \qquad \sigma=\sqrt{1-\ft{q^2}{\rho^2}},
\qquad \rho\ge q.
\eea
 Without loss of generality, we choose a constant time-slice and set $\theta=\ft{\pi}{2}$. The wormhole metric \eqref{solution2} and \eqref{theo2} becomes
\be
ds^2=(\sigma^2h_\pm)^{-1}d\rho^2+\rho^2 d\psi^2.
\ee
Embedding this hypersurfce to the three dimensional Euclidean space
\be
ds^2=dz^2+d\rho^2+\rho^2 d\psi^2,
\ee
we have
\be\label{dzdr}
\ft{dz}{d\rho}=\pm \sqrt{\ft{1}{\sigma^2h_\pm}-1}.
\ee
The $z$-$\rho$ relations of the two parts of spacetime are determined by
\be\label{zz}
\ft{dz_+}{d\rho}=\sqrt{\ft{1}{\sigma^2h_+}-1},\qquad \ft{dz_-}{d\rho}=-\sqrt{{\ft{1}{\sigma^2h_-}-1}},
\ee
where $z_+$ and $z_-$ take the ranges of $(0,+\infty)$ and $(-\infty,0)$ respectively. Integrating \eqref{zz} we get the embedding diagrams. The three embedding diagrams with different charge are shown in Fig \ref{fig:wh2}.  The diagrams of $z=z(\rho)$ clearly show that the two spacetimes connected by the wormhole are different.

Now we turn to investigate the flare-out condition of the wormholes. For a generic spherically symmetric wormhole metric, i.e,
\be\label{ds2}
ds^2 = - h(\rho) dt^2 + \fft{d\rho^2}{\sigma^2h(\rho)} + \rho^2 d\Omega_{2}^2,
\ee
there is a flare-out condition in the sense of the geometry is the minimality of the wormhole throat in the embedded spacetime. The condition is given by
\be
\ft{d\rho}{dz}|_{\rho=\rho_t}=0,\qquad\qquad
\ft{d^2\rho}{dz^2}|_{\rho=\rho_t}>0,
\ee
where $\rho_t$ denotes the location of wormhole throat. It means that
\be
\ft{d^2\rho}{dz^2}|_{\rho=\rho_t}=\ft{d\rho}{dz}\ft{d}{d\rho}\ft{d\rho}{dz}|_{\rho=\rho_t}=\ft{\sigma}{2}\ft{\sigma h'+2h \sigma'}{(h\sigma^2-1)^2}|_{\rho=\rho_t}>0,
\ee
or
\be
(\sigma h'+2h \sigma')|_{\rho=\rho_t}>0.
\ee
For our wormhole solution \eqref{solution}, this requires
\be\label{flareout}
\ft{4q^2+\gamma_1Q^2}{2q^3}>0.
\ee
Intriguingly, the condition is independent of $\gamma_2$.

\section{Black hole without singularities}\label{sec4}

\subsection{Black hole solution}

The solution \eqref{ax} can also describe a black hole if there exists an event horizon. An event horizon is specified by $r=r_h$ where $g_{tt}=-h=0$. It's easy to verify this is nothing but
\be\label{bhcon}
 Q^2\geq\ft{8q^2(\gamma_1+\sqrt{\gamma_1^2+\gamma_2^2})}{\gamma_2^2},
\ee
or in terms of $(Q_e, M)$,
\be
(\ft{Q_e}{\gamma_1})^2\leq\ft{8M^2(-\gamma_1+\sqrt{\gamma_1^2+\gamma_2^2})}{\gamma_2^2}\,.
\ee
The inequality is saturated by the extremal black hole.

The equation $h(r)=0$ determines the Killing horizon's location. We find there exist two roots of this equation,
 \begin{equation}\label{rmp}
   r_{\pm}=\frac{\gamma_2 Q^2}{8q}\pm\sqrt{\frac{\gamma_2^2 Q^4}{64q^2}-q^2-\frac{\gamma_1Q^2}{4}}.
 \end{equation}
It is not a usual black hole which has a singularity inside. It follows from \eqref{cur} that the black hole solution \eqref{ax} has no curvature singularity unless $q=0$. The typical singularity associated with black holes is now replaced by a wormhole throat or a bounce at $r=0$. Thus a particle in the $r>0$ region that falls into the horizon will tunnel to another universe in the $r<0$ region. We denote this two cases by:

(1) Type I black hole: Both horizons are located in the $r>0$ region. It is analogous to the Reissner-Nordstr\"om (RN) black hole, with one outer and one inner horizon.  The essential difference is that the curvature singularity at $r=0$ of the RN black hole is now replaced by a wormhole.

(2) Type II black hole: The two horizons are located both sides of $r=0$. The configuration can be viewed as a pair of black hole and white hole, and the ``wormhole throat'' now becomes a bounce from a black hole to a white hole\cite{Simpson:2018tsi}.

\subsection{Black hole thermodynamics}

We first study the black hole thermodynamics. We first consider the type I black hole viewed from the $r>0$ region.  In this case, the analysis is similar to the usual RN black hole.  The surface gravity is $\kappa=\ft{h'}{2}$ in the metric like \eqref{ax}. And we obtain the surface gravity for the two horizons in terms of $r_+$ and $r_-$ are given by
 \begin{equation}
   \kappa_{+}=\frac{r_{+}-r_{-}}{2 (r_{+}^2+q^2)}  \;,\qquad \kappa_{-}=\frac{r_{-}-r_{+}}{2 (r_{-}^2+q^2)}.
 \end{equation}
 And the temperature $T=\ft{\kappa}{2\pi}$, so
 \be
 T_{+}=\frac{r_{+}-r_{-}}{4\pi (r_{+}^2+q^2)}  \;,\qquad T_{-}=\frac{r_{-}-r_{+}}{4\pi (r_{-}^2+q^2)}.
 \ee
Note that here we follow the same view as in \cite{Huang:2016fks,Cvetic:2018dqf} that the inner horizon has negative temperature. The entropy of the black hole is related to the horizon area, which are given by
  \begin{equation}
   S_{+}=\pi  (r_{+}^2+q^2)  \;,\; S_{-}=\pi  (r_{-}^2+q^2).
 \end{equation}
Obviously, there is an interesting relation that
   \begin{equation}
        T_{+}S_{+}+ T_{-}S_{-}=0 \label{TSsum}.
   \end{equation}
As was pointed out in Ref \cite{Cvetic:2018dqf}, this is also true for the RN black hole.  This vanishing sum relates to the fact that the entropy product depends only on $Q_e$.  In our case, we indeed find
   \begin{equation}
      S_{+} S_{-}=\frac{\pi^2(\gamma_1+\gamma_2) Q_e^2}{16\gamma_1^2}.
   \end{equation}
The electric potentials at both horizons are $\Phi_\pm = \int_{r_\pm}^0 \xi' dr$ and we find that they are
    \begin{equation}
        \Phi_{+}=\frac{Q}{4\gamma_1} \frac{\gamma_1 r_{+}+\gamma_2}{r_{+}^2+q^2}  \;,\; \Phi_{-}=\frac{Q}{4\gamma_1} \frac{\gamma_1 r_{-}+\gamma_2}{r_{-}^2+q^2}.
   \end{equation}
It is then straightforward to verify that the first law of the black hole thermodynamics satisfied at both horizons, namely
    \begin{equation}
    \begin{split}
        \delta M &= T_+\delta S_{+}+\Phi_{+}\delta Q_{e}  \\
                      &= T_-\delta S_{-}+\Phi_{-}\delta Q_{e}.
   \end{split}  \label{BHthermo}
   \end{equation}
These two equivalent expressions of the first law of black hole thermodynamics and \eqref{TSsum} imply that
   \begin{equation}
    \begin{split}
        \delta (S_+S_-) &= \frac{(T_{+}S_{+}+ T_{-}S_{-})}{T_{+}T_{-}} \delta M
                                       -(\frac{\Phi_{+}S_{-}}{T_{+}}+\frac{\Phi_{-}S_{+}}{T_{-}})  \delta Q_{e} \\
                      &= -(\frac{\Phi_{+}S_{-}}{T_{+}}+\frac{\Phi_{-}S_{+}}{T_{-}})  \delta Q_{e}.
   \end{split}
   \end{equation}
It means the mass does not appear in entropy product $S_{+}S_{-}$ when $S_{+}$ and $S_{-}$ are treated as functions of mass and charge.

For the type II black holes, the first law becomes
\be
\delta M_+ = T_+\delta S_{+}+\Phi_{+}\delta Q_{e}\,,\qquad
\delta M_- = T_-\delta S_{-}+\Phi_{-}\delta Q_{e}\,,
\ee
with the understanding that $(M_-, T_-, \Phi_-)$ changes the sign of those obtained from the view at $r>0$ region.  With this subtlety understood, the conclusions of the type I black holes can carry over to the type II as well.

\section{Causal structure}\label{cs}

The solutions of wormhole and black hole we constructed in the previous sections have many different kinds of causal structure. To analyse them, we would like to reduce the range of the parameter space first.
The sign of $\gamma_2$ does not affect the spacetime structure since we can simply redefine the coordinate $r\rightarrow -r$ to absorb the sign. We can therefore focus on the parameter range $\gamma_2>0$.
On the other hand, the sign of $\gamma_1$ yields inequivalent situations. If it is negative, the hypersurface $r=0$ where the sphere area is minimum can be timelike, spacelike or null. Thus, we classify three probabilities by the sign of $h(0)$. The expression of $h(0)$ is
   \begin{equation}\label{h0}
    h(0)=1+\frac{\gamma_1 Q^2}{4q^2}.
   \end{equation}
 The expression of $r_{\pm}$ can be simplified as
   \begin{equation}
    r_{\pm}=M\pm\sqrt{M^2-q^2h(0)}.
   \end{equation}

 \begin{figure}[h]
	\centering
	\includegraphics[width=4.5cm]{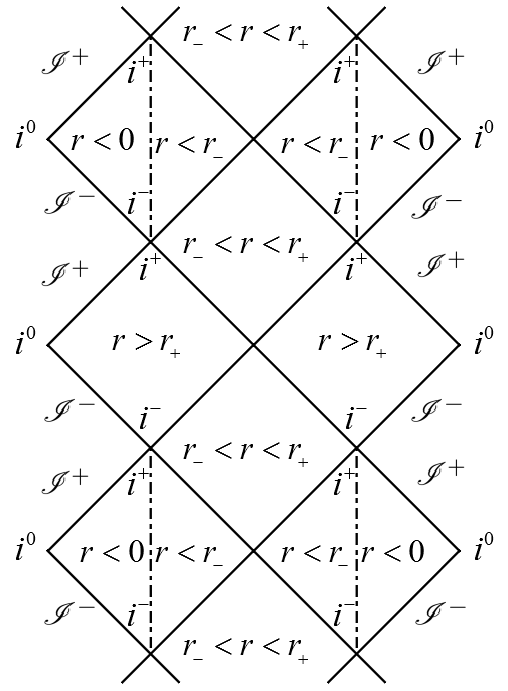}
	\includegraphics[width=4cm]{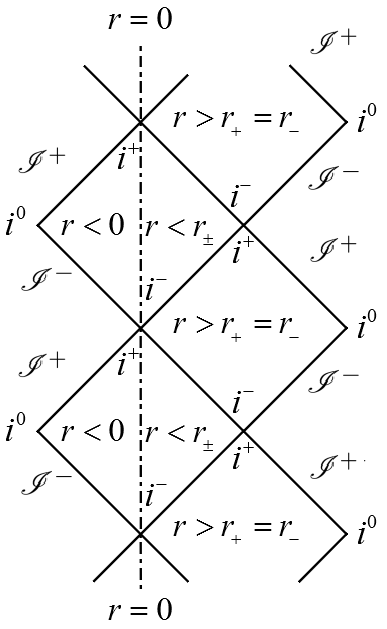}
	\includegraphics[width=4cm]{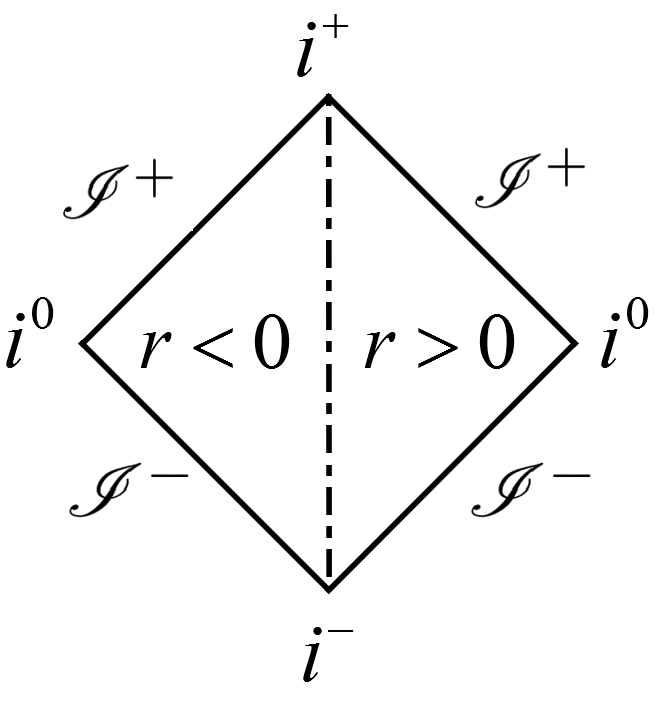}
	\caption{The three diagrams are all satisfied $h(0)>0$. The first diagram shows the type I black hole, namely $M^2>q^2h(0)$. The second diagram shows the extreme type I black hole, namely $M^2=q^2h(0)$. The last one shows the traversable wormhole, namely $M^2<q^2h(0)$.}
	\label{fig:hgeq}
\end{figure}
 In the case of $h(0)>0$, which can be achieved when $\gamma_1>0$ , or when $\gamma_1<0$, but $q^2>\frac{-\gamma_1 Q^2}{4}$. In this case, the hypersurface r=0 is definitely timelike. We draw the Penrose-Carter diagrams for the situations of two non-extreme horizons (type I black hole), extreme horizon (extreme type I black hole) and no horizon (wormhole) in Figure \ref{fig:hgeq}.
 The causal structure is just like a RN black hole but with its timelike singularity replaced by a travelable wormhole throat, or a Kerr black hole if one treats the disk surrounded by the singular ring as a squeezed wormhole throat. This throat connects another asymptotically flat region which looks like a RN black hole with negative mass.
 \begin{figure}[h]
	\centering
	\includegraphics[width=4cm]{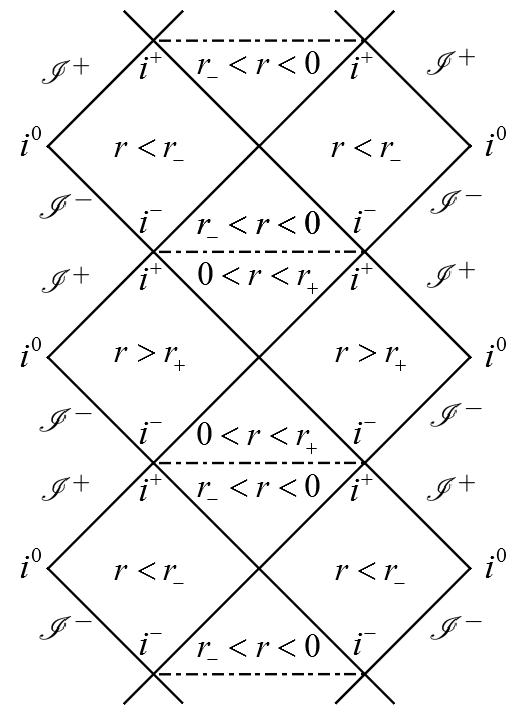}
	\includegraphics[width=4cm]{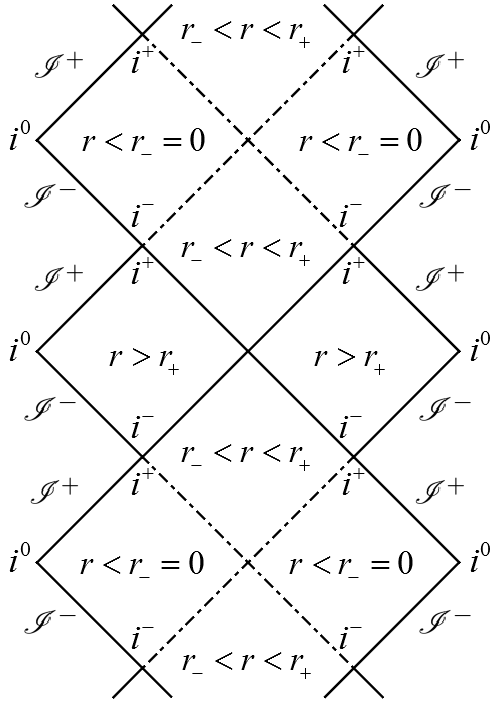}
	\includegraphics[width=4cm]{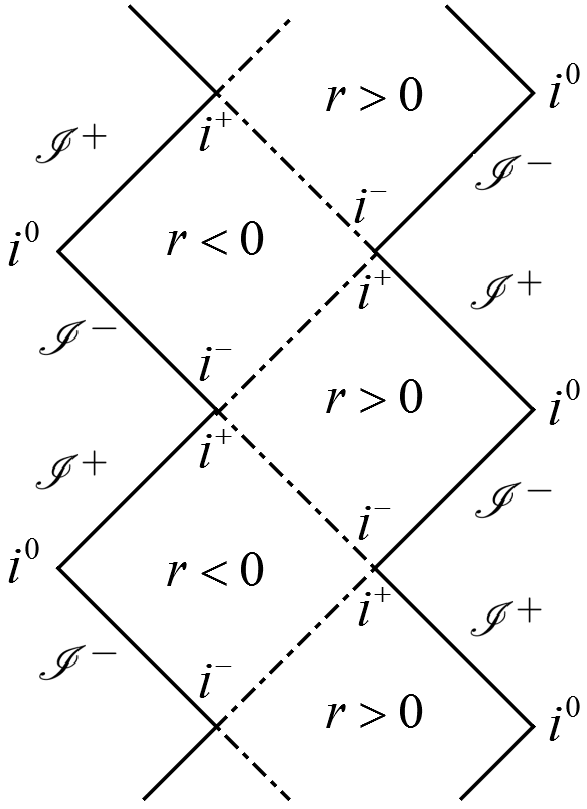}
	\caption{The left one shows the type II black hole, namely $h(0)<0$. The middle one shows the critical situation of type II black hole with the bounce occurs at $r=0$, which means $h(0)=0$. The last one corresponds the extreme type I black hole with $\gamma_2=0$, which is similar with the case of $a=2m$ in Ref\cite{Ashtekar:2018lag}.}
	\label{fig:hleq}
\end{figure}

In the case of $h(0)<0$, namely type II black hole, the hypersurface $r=0$ is spacelike so $r=0$ is not a travelable throat. Note that in this case we have $\gamma_1<0$, which leads to the phantom-like electromagnetic field at the asymptotic region according to \eqref{inZ}. The Penrose-Carter diagram was drawn in the left plot of Figure \ref{fig:hleq}.
 The hypersurface connects a trapped region with an anti-trapped region, namely it can be regarded as the tunneling from a black hole to a white hole. The idea of tunneling from black hole to white hole has been constructed in semi-classical limit of Loop Quantum Gravity and was expected to contribute some ideas to solve black hole information paradox\cite{Ashtekar:2018lag}.

The metric structure of this kind of spacetime is analogous to \eqref{viss}, which was described as ``black bounce,'' in Ref \cite{Simpson:2018tsi,Simpson:2019cer}. However, there was hitherto no concrete theory that could give rise to such solutions. In this paper, we showed that the black bounces can arise as exact solutions in our theory. In particular, when $\gamma_2=0$ and $\gamma_1<0$, the metric in (\ref{ax}) becomes symmetric black bounce between $r>0$ and $r<0$, namely
 \be
 ds^2=-(1+\ft{\gamma_1 Q^2}{4(r^2+q^2)})dt^2+(1+\ft{\gamma_1 Q^2}{4(r^2+q^2)})^{-1}dr^2+(r^2+q^2)d\Omega_2^2.
 \ee
In this case, the mass becomes zero. For simplification, we consider without loss of generality that $\gamma_1=-1$.  When $2q<Q$, the solution describes a black bounce; when $Q=2q$, the solution is a one-way wormhole; when $2q>Q$, the solution is a traversable wormhole.  These correspond exactly to the
$a<2m$, $a=2m$ and $a>2m$ of the metric \eqref{viss} respectively.

For general solutions, the spacetime characteristics can be solely determined by the sign choice of $h(0)$ given in \eqref{h0}. The Penrose-Carter diagrams for $h(0)>0$ are shown in Figure \ref{fig:hgeq} and the ones with $h(0)< 0$ are shown in Figure \ref{fig:hleq}.  These Penrose diagrams are analogous to those in the Ref \cite{Simpson:2018tsi}. The case of $h(0)=0$ can be treated as the critical situation that transits between a travelable wormhole and to a black bounce. The Penrose-Carter diagram is given as the right plot of Figure \ref{fig:hleq} and it shows that the bounce occurs on the null hypersurface $r=0$.

\section{Shadow of wormhole and black hole}\label{sec5}

An interesting feature of black holes or wormholes is that they attracts massless particles such as photons which can form photon spheres. The unstable photon orbits corresponding to logarithmical divergence of the deflection angle. The photon spheres can produce shadows of the objects, such that the area of the object we detect is bigger than it truly is.

\subsection{General discussions}

For the metric like \eqref{ax}, i.e,
\be
ds^2=-h dt^2+(h)^{-1}dr^2+(r^2+q^2)d\Omega_{2}^2.
\ee
Suppose there is a particle moving in the spacetime. Without loss generality, we restrict the motion to lie in the $\theta=\ft{\pi}{2}$ plane. The lagrangian for the particle is given by\cite{Goulart:2017iko}
\be
2{\cal L}=-h\dot{t}^2+\ft{1}{h}\dot{r}^2+(r^2+q^2)\dot{\psi}^2,
\ee
where a dot represent a derivative with respect to the affine parameter. There are two first-integral quantities namely the energy and angular momentum, given by
\be
E=-\ft{\partial{\cal L}}{\partial \dot{t}}=h\dot{t},\qquad
L=\ft{\partial{\cal L}}{\partial \dot{\psi}}=(r^2+q^2)\dot{\psi}
\ee
respectively. The geodesic condition for the particle $\kappa=g_{\mu\nu}\dot{x}^\mu\dot{x}^\nu$ can be written as
\be\label{dotr}
\dot{r}^2+V_{eff}=E^2,\qquad V_{eff}=-h\kappa+\ft{h}{r^2+q^2}L^2,
\ee
where $\kappa=-1$ for massive particle and $\kappa=0$ for massless particle.

We only consider massless particle like photon in this section; therefore, we set $\kappa=0$ in the following discussion. In the case, from Ref \cite{Bozza:2002zj} we know that
\be
\ft{d\psi}{dr}= \ft{1}{(r^2+q^2)\sqrt{\ft{1}{b^2}-\ft{h}{r^2+q^2}}},
\ee
where $b=\ft{L}{E}$ is the impact parameter which relates to the closest approach distance. For the turning point $r_0$ where $\ft{dr}{d\psi}=0$ gives the relation
\be
b=\sqrt{\ft{r_0^2+q^2}{h}}.
\ee
It is easy to obtain the deflection angle,
\be
\alpha_{del}=2\int_{r_0}^{\infty}d\psi
=2\int_{r_0}^{\infty}{\ft{1}{(r^2+q^2)\sqrt{\ft{1}{b^2}-\ft{h}{r^2+q^2}}}}dr-\pi.
\ee
In the sense of mathematics, an orbit is so called ``photon sphere" if it satisfies
\be\label{condition}
\dot{r}=0,\qquad\qquad \ddot{r}=0\qquad\qquad \text{and}\qquad \dddot{r}>0,
\ee
which makes the logarithmical divergence of the deflection angle $\alpha_{del}$. Now we discuss it for wormhole and black hole as follow.

\subsection{Photon sphere}

To begin with, we rewrite \eqref{dotr} as
\be
\dot{r}^2=(E^2-V_{eff}).
\ee
Then the condition \eqref{condition} becomes
\be\label{con}
V_{eff}(r_0)=E^2,\qquad \ft{\partial V_{eff}}{\partial r}|_{r=r_0}=0,\qquad
\ft{\partial^2 V_{eff}}{\partial r^2}|_{r=r_0}<0.
\ee
\begin{figure}[h]
	\centering
	\includegraphics[width=7cm]{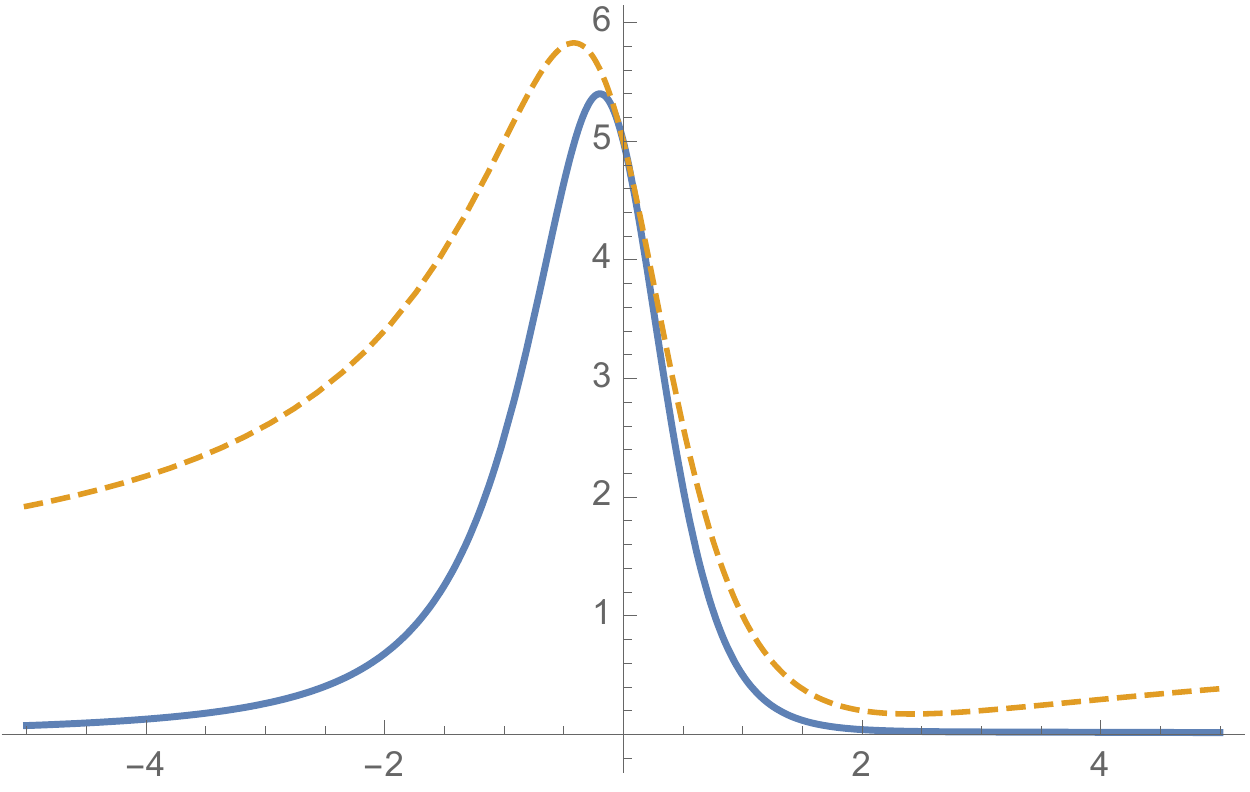}
	\includegraphics[width=7cm]{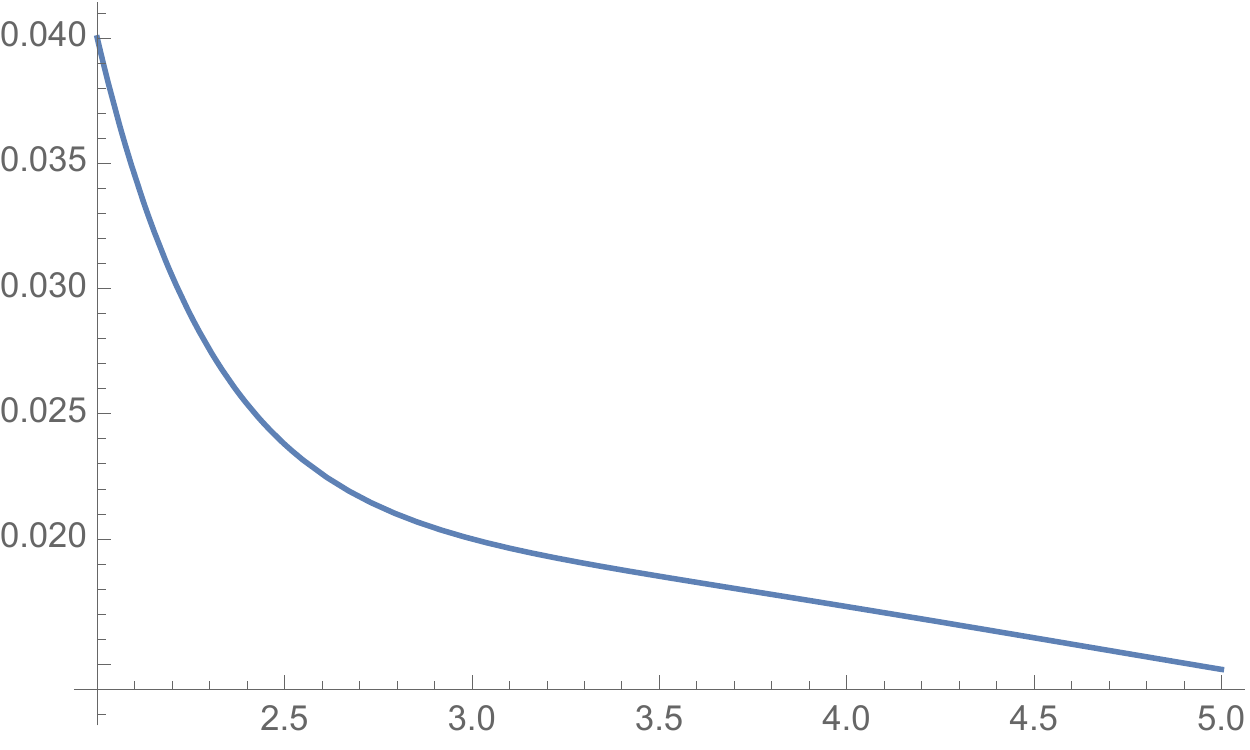}
	\caption{For the case of one photon sphere wormhole, we show the $V_{eff}$ is depicted by the solid line and $h$ is depicted by the dashed line. The right one is the details of $V_{eff}$ and it shows that there is no photon sphere in the range $r>0$. We have chosen $L=1,\gamma_1=1,\gamma_2=1,q=1,Q=4$.}
	\label{fig:vwh1}
\end{figure}
\begin{figure}[h]
	\centering
	\includegraphics[width=7cm]{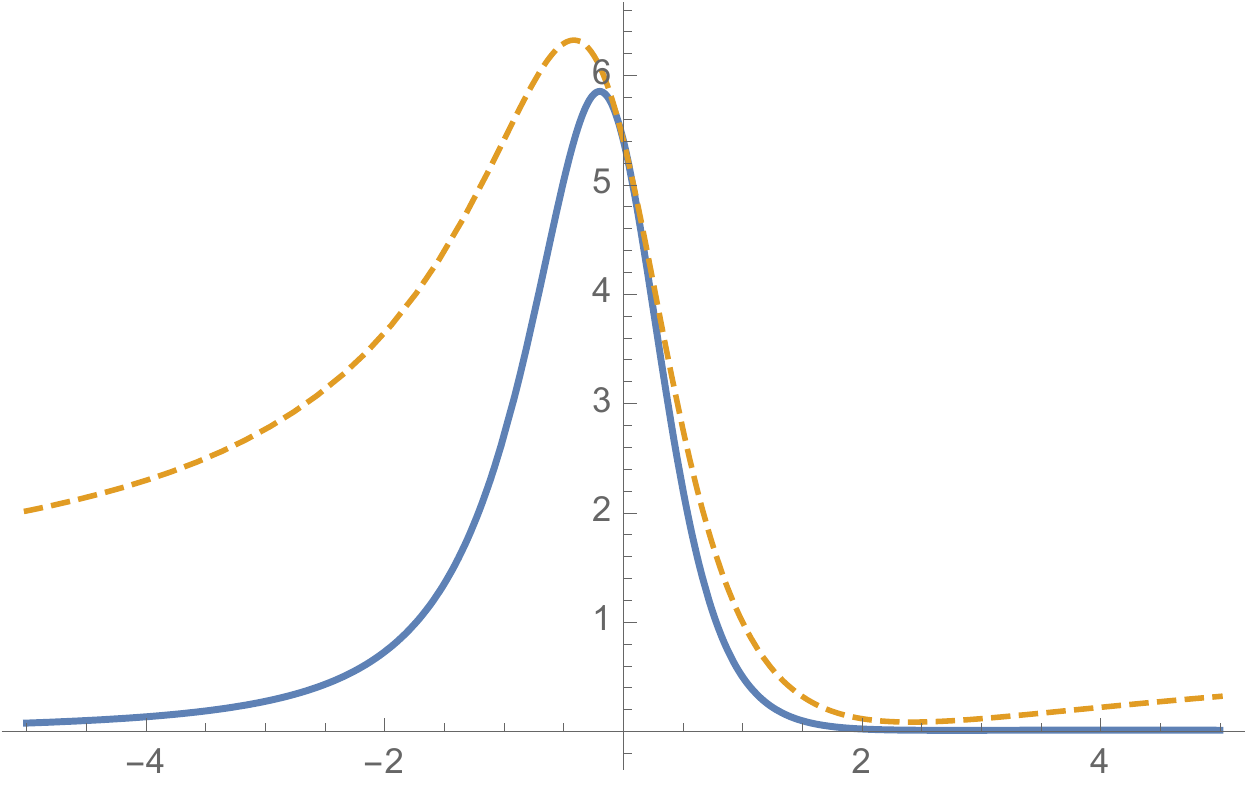}
	\includegraphics[width=7cm]{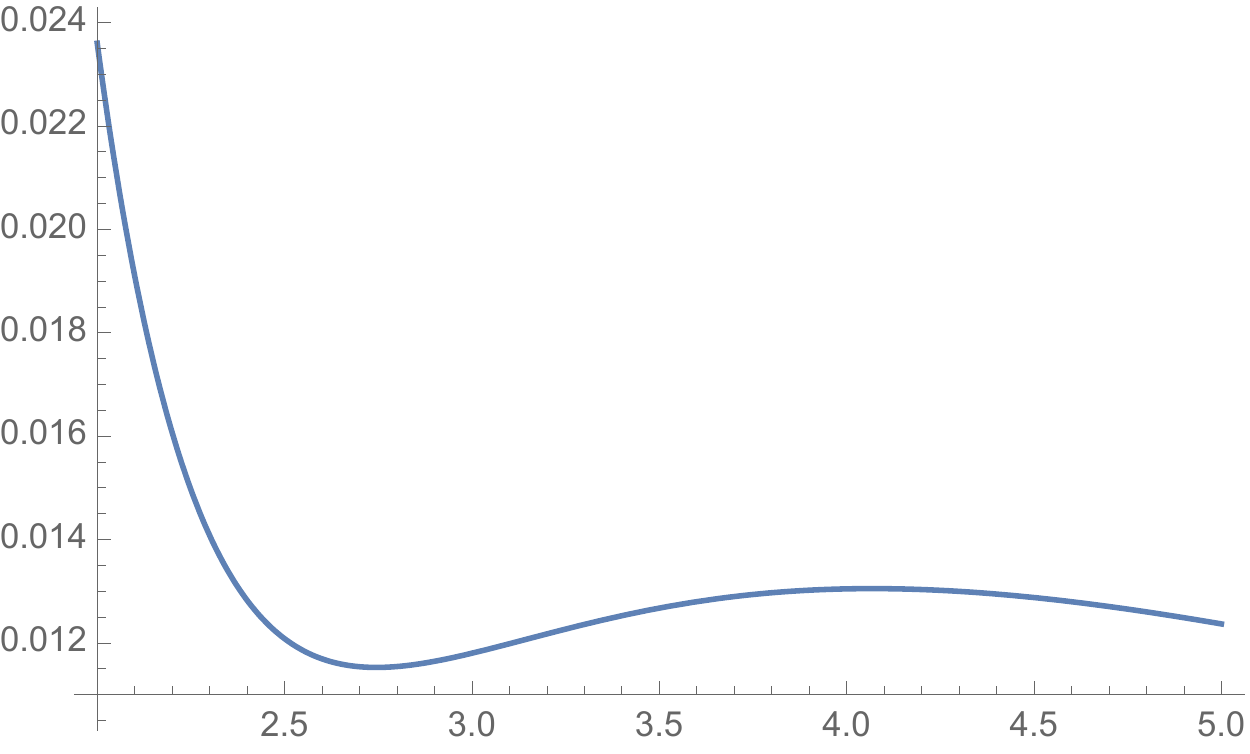}
	\caption{For the case of two photon sphere wormhole, we show the $V_{eff}$ is depicted by the solid line and $h$ is depicted by the dashed line. The right one is the details of $V_{eff}$ and it shows that there is a photon sphere in the range $r>0$. We have chosen $L=1,\gamma_1=1,\gamma_2=1,q=1,Q=4.2$.}
	\label{fig:vwh2}
\end{figure}
\begin{figure}[h]
	\centering
	\includegraphics[width=7cm]{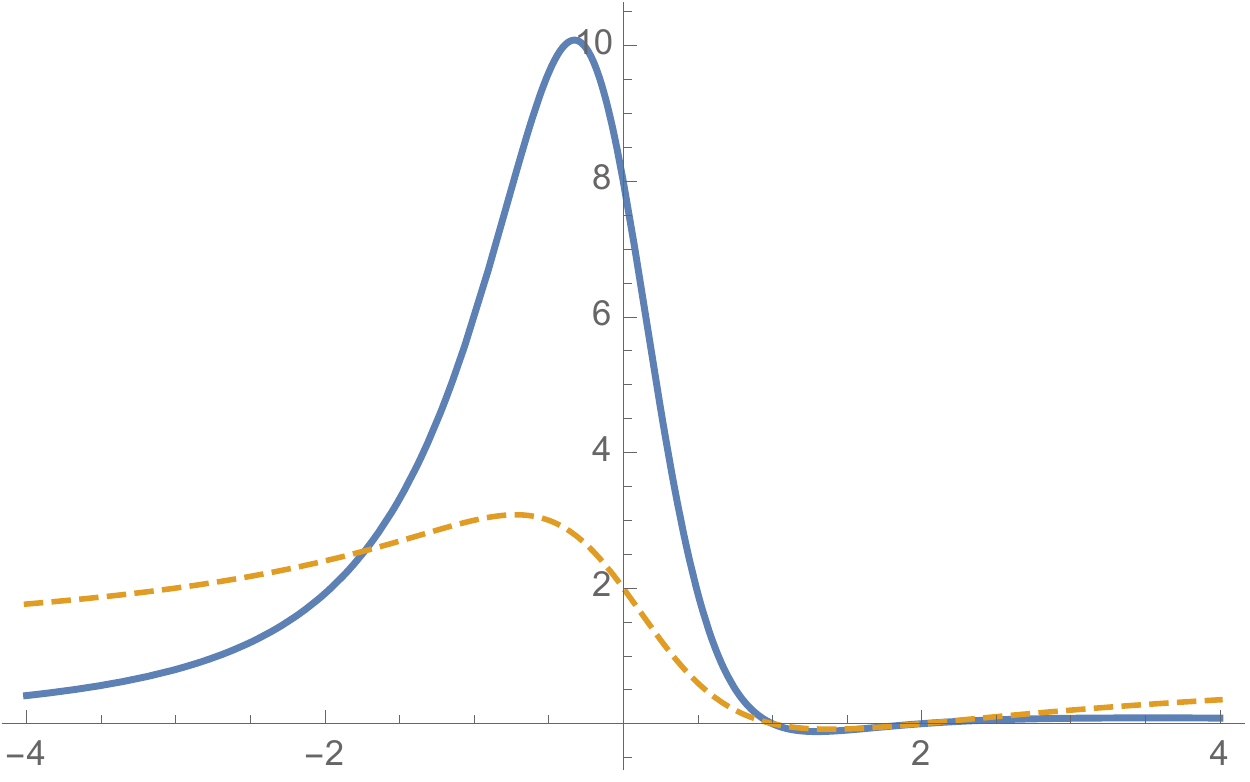}
	\includegraphics[width=7cm]{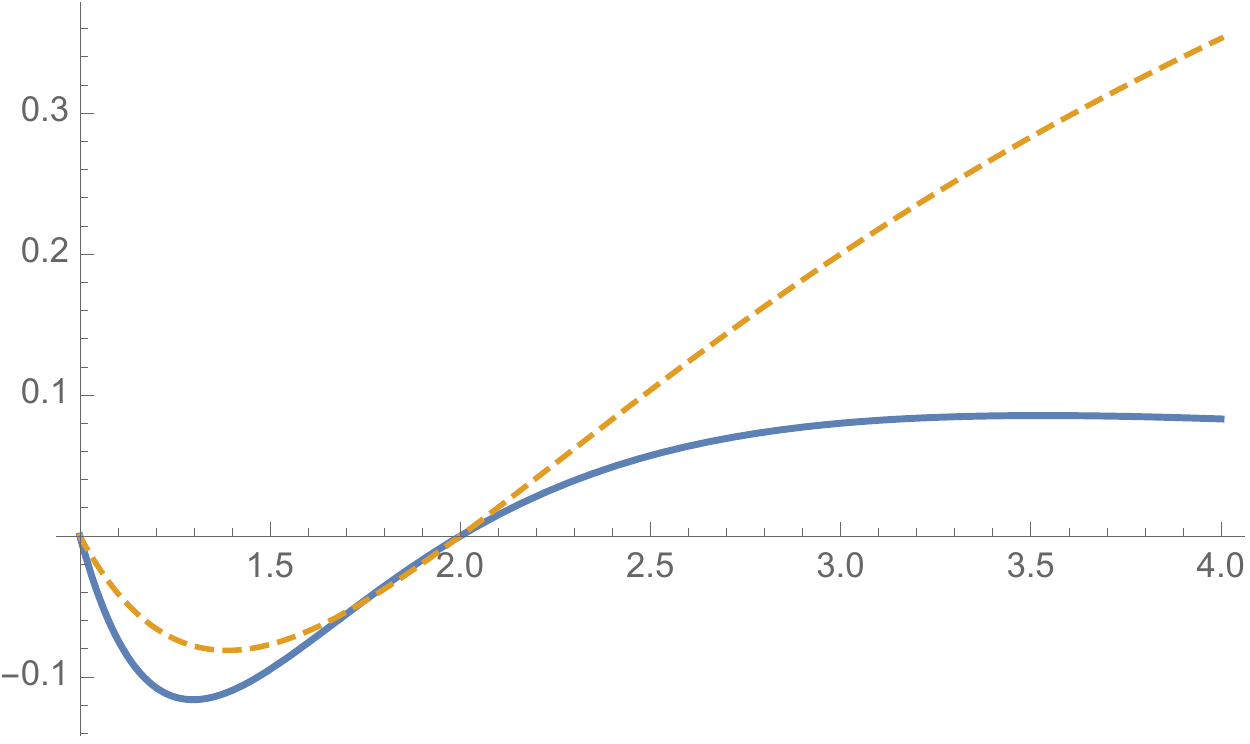}
	\caption{For the case of type I black hole, we show the $V_{eff}$ is depicted by the solid line and $h$ is depicted by the dashed line. The left one shows there is a photon sphere at the outsides of the inner horizon. The right one is the details of $V_{eff}$ and it shows that there is a photon sphere at the outsides of the outer horizon. We have chosen $L=2, r_-=1, r_+=2, q=1$.}
	\label{fig:vbh1}
\end{figure}
\begin{figure}[h]
	\centering
	\includegraphics[width=7cm]{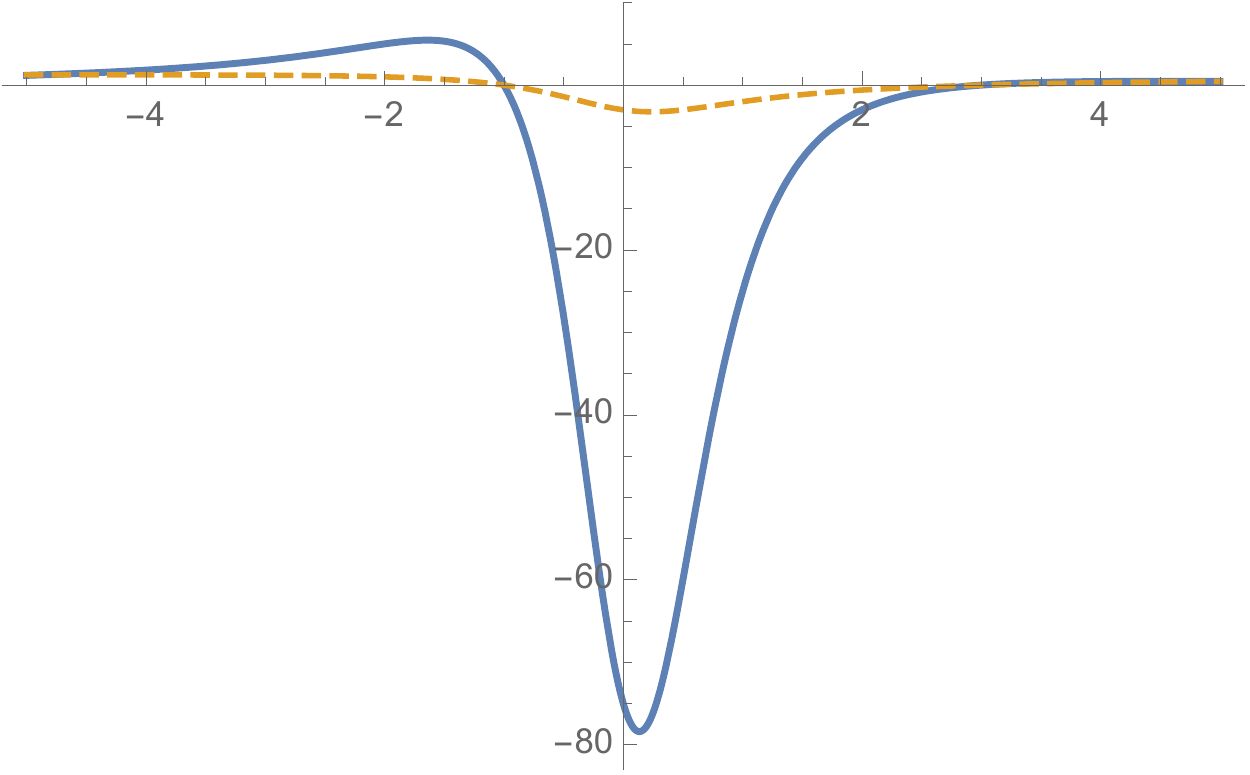}
	\includegraphics[width=7cm]{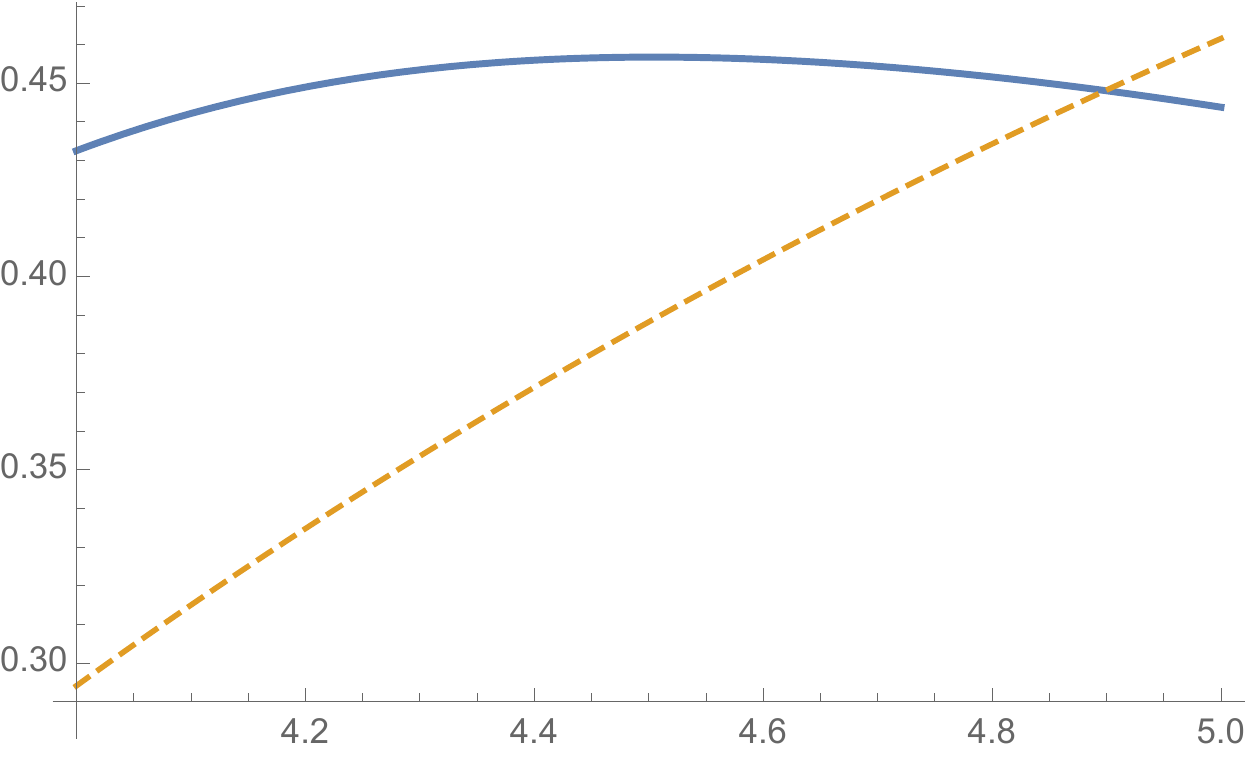}
	\caption{For the case of type II black hole, we show the $V_{eff}$ is depicted by the solid line and $h$ is depicted by the dashed line. The left one shows there is a photon sphere at the outsides of the white hole horizon. The right one is the details of $V_{eff}$ and it shows that there is a photon sphere at the outsides of the black hole horizon. We chose $L=5, r_-=-1, r_+=3, q=1$.}
	\label{fig:vbh2}
\end{figure}

It is worth pointing out that photon sphere corresponds to the maximum of the effective potential $V_{eff}$. Substituting the explicit form of $V_{eff}$ into \eqref{con} gives
\bea
&&\ft{h L^2}{r^2+q^2}=E^2,\\
&&-2r h+(q^2+r^2)h'=0,\label{con22}\\
&&-2(q^2-3r^2)h+(q^2+r^2)\big(-4r h'+(q^2+r^2)h''\big)<0.\label{con2}
\eea
For a photon with energy $E$ and angular momentum $L$, \eqref{con22} can tell us where the photon spheres are. Thus we substitute the solution \eqref{ax} into \eqref{con22}, which reduces to a cubic equation
\be\label{cubic}
\ft{\partial V_{eff}}{\partial r}=8q r^3-3\gamma_2 Q^2 r^2+(8q^3+4q \gamma_1 Q^2)r+\gamma_2q^2Q^2=0.
\ee
The equation thus has either a single real root or three real roots.

Now we are going to prove that there is always a real root of \eqref{cubic} which plays the role of photon sphere. In the limit $r\to-\infty$, the behavior of $V_{eff}$ is given by
\be
V_{eff}\sim\ft{L^2}{r^2},
\ee
which indicates that $V_{eff}\to+0$ while $r\to-\infty$. If we take a Taylor expansion at $r=0$, it gives
\be \label{tay}
V_{eff}=-\ft{L^2(4q^2+\gamma_1 Q^2)}{4q^4}-\ft{L^2 Q^2 \gamma_2}{4q^5}r+{\cal{O}} (r^2).
\ee
Since all the coefficients of the sub-leading term of \eqref{tay} are positive, it shows that
\be
V_{eff}(0)<V_{eff}(0-\epsilon),
\ee
where $\epsilon$ is a positive infinitesimal. According to the mean value theorem, there is at least a maximum of $V_{eff}$ in the range $(-\infty,0)$ which corresponds to the location of the photon sphere. For the wormhole solution, it may has this only photon sphere. Of course, it could also has two photon spheres of wormhole if there are two roots and one is a photon sphere while another is a stable photon orbit.

The case of black hole is different.  For convention, we rewrite the $h$ for the black hole solution as
\be
h=\ft{(r-r_+)(r-r_-)}{r^2+q^2},
\ee
where $r_+$ and $r_-$ are defined by \eqref{rmp}. Recall the definition of $V_{eff}$, we can also rewrite it as
\be
V_{eff}=\ft{(r-r_+)(r-r_-)}{(r^2+q^2)^2}L^2.
\ee
Now we would like to prove that there are always two photon spheres of the black holes located at the outsides of the horizons. To begin with, consider the value of $V_{eff}$ in the region $(-\infty,r_-)$, the value is plus and tends to $0$ while $r$ tends to $r_-$ or $-\infty$. Applied the mean value theorem again, it is shown that there is at least a photon sphere in the region.
Then we consider the region $(r_+, \infty )$. The asymptotic behavior of $V_{eff}$ at $\infty$ is given by
\be
V_{eff}(\infty)\sim\ft{L^2}{r^2},
\ee
which means that $V_{eff}(\infty)=+0$. Note that $V_{eff}(r_+)=0$, and hence the  mean value theorem ensures there is a maximum which is a photon sphere located outsides the horizon.

\section{Conclusion}\label{sec6}

In this paper, we considered a class of EMS theories, with a phantom scalar field that is non-minimally coupled to the Maxwell field. Generalizing the Ellis wormhole, we constructed new exact electrically charged traversable wormholes. We found that, in our wormhole solutions, the electric charge $Q_e$ enters the solution only through the metric function $h$. Compared to the various constructions in literature, this may be the simplest way to charge the Ellis wormhole. The charging also brings mass to the new wormhole solutions. While Ellis wormhole is massless and symmetric, the inclusion of charges leads to asymmetric wormholes, and its mass is different viewed from the two different asymptotic regions. We drew the embedding diagrams of these wormholes.

We found that black hole with two horizons emerge as the value of charges and hence the mass increases. A novel feature is that these black holes have no curvature singularity. We found that there existed two types of black holes. The type I is analogous to the RN black hole but with the singularity replaced by a wormhole throat. The type II is a ``black hole-white hole pair" connected by a bounce; it is a concrete example of the recently proposed black bounce. We also discussed the black hole thermodynamics of these solutions.
The causal structures of wormholes and black holes are very different.  We considered each situation and drew the Carter-Penrose diagrams for the solutions.

Finally, we examined the gravitational lensing effect by the wormhole and black hole. We computed their photon spheres. The wormholes can either have only one photon sphere, or two with one in each side of wormhole throats. However, in the black hole case, there are always two photon spheres, with one outside each horizon. The photon spheres can produce the shadows of the wormhole or black hole, which may be valuable for observation.

For future work, we expect to study the uniqueness theorem of our solutions like \cite{Lazov:2017tjs} for previous developed charged wormholes. And, we would like to consider dynamic wormhole or black hole solutions. Another direction worth further exploring is to study the asymptotic (A)dS wormhole in analogous EMS theories but with a scalar potential. We expect it would shed some light on the study of holographic models and cosmology.

\section*{Acknowledgement}

We are grateful to  Hong L\"u for  useful  discussions and proofreading the manuscript. J.Y. is supported by the China Scholarship Council and the Japanese Government (Monbukagakusho-MEXT) scholarship.  H.H. is supported in part by NSFC Grants No. 11875200 and No. 11475024. H.H.~is grateful to the Center for Joint Quantum Studies, Tianjin University, for hospitality, during the course of this work.

\appendix
\setcounter{equation}{0}

\section{Constructing the theory and solutions}

In this paper, we consider a general class of EMS theory with the following Lagrangian,
\bea\label{Ztheory}
&&\mathcal{L}=\sqrt{-g}(R+\ft{1}{2}(\partial \phi)^2-\ft{1}{4}Z^{-1}F^2),\qquad Z=Z(\phi).
\eea
The theory admits the Ellis wormhole when the Maxwell field is turned off.  In this appendix, we would like to generalize the Ellis wormhole by including the charge. Some of the generalization was given in Ref.\cite{Goulart:2017iko}, where the scalar field is modified by the new integration constant from the Maxwell equation.  In this appendix, we would like to find the coupling function $Z(\phi)$ such that the theory admits exact charged wormhole solutions but with the scalar remaining identically the same as the (neutral) Ellis wormhole. The benefit is that the solution becomes simpler than those in Ref.\cite{Goulart:2017iko}.

The equations of motion associated with the variation of the scalar field $\phi$, the Maxwell field $A_\mu$ and the metric $g_{\mu\nu}$ are respectively given by
\bea\label{eom}
&&\Box\phi =\fft 14 \fft{\partial Z^{-1}}{\partial \phi}F^2 \,,\cr
&&\nabla_\mu\big(\sqrt{-g}Z^{-1}F^{\mu\nu}\big) = 0\,, \cr
&&E_{\mu\nu} \equiv R_{\mu\nu}-\ft 12 R g_{\mu\nu}-T_{\mu\nu}^{A}-T_{\mu\nu}^{\phi}=0\,,
\eea
with
\bea
T_{\mu\nu}^{A} &=&\ft 12 Z^{-1}\Big(F_{\mu\nu}^2-\ft 14 g_{\mu\nu} F^2 \Big) \,,\nn\\
T_{\mu\nu}^{\phi} &=&-\ft 12\partial_\mu\phi \partial_\nu \phi+\ft 14 g_{\mu\nu}(\partial \phi)^2\,.\label{TaTphi}
\eea
The ansatz for the most general static and spherically symmetric electrically-charged solution is
\be\label{ds}
ds^2=-h dt^2+(\sigma^2 h)^{-1}d\rho^2+\rho^2 d\Omega_{2}^2,\quad \phi=\phi(\rho), \quad A=\xi(\rho)dt,
\ee
where the metric functions depend only on $\rho$, namely $h=h(\rho)$ and $\sigma=\sigma(\rho)$.

The Maxwell equation can be solved straightforwardly, given by
\be
\xi'=\ft{QZ}{\sigma \rho^2},
\ee
where the prime is a derivative with respect to $\rho$. Substitution of \eqref{ds} into the Einstein equation \eqref{eom}, we get
\bea\label{eomm}
E_0^0=0:&&-4\rho^2+ZQ^2+4\rho^3\sigma^2 h'-\rho^2h\sigma\big(\sigma(4-\rho^2\phi'^2)+8\rho \sigma'\big)=0, \label{eom00}\\
E_1^1=0:&&-4\rho^2+ZQ^2+4\rho^3\sigma^2 h'-\rho^2h\sigma(4+\rho^2\phi'^2)=0, \label{eom11}\\
E_i^i=0:&&-ZQ^2+2\rho^3\sigma\sigma'(2h+\rho h')-\rho^3\sigma^2(4h'-\rho h\phi'^2+2\rho h'')=0\label{eom22}.
\eea
It is a standard trick to consider the linear combination of the equations $E^0_0-E^1_1=0$, which simplifies dramatically\cite{Feng:2013tza}:
\be\label{sigma}
\rho\sigma\phi'^2-4\sigma'=0.
\ee
As was discussed earlier, we require that $\phi$ is independent on $Q$, it follows that $\sigma$ is also independent of $Q$, since the integration constant can be absorbed into $h$. Note that the Einstein equations are invariant and the scalar equation is covariant under $\rho\rightarrow -\rho$. This implies that for fixed integration constant, the solution may not be invariant under $\rho\rightarrow -\rho$, and hence asymmetric wormholes can arise.

It should be emphasized that the equation (\ref{sigma}) is identically the same as that for solving the Ellis wormhole for which the charge is turned off.  As we discussed earlier, we shall keep this part of the solution the same as the Ellis wormhole, thus, we have
\be\label{phi}
\phi=2\arcsin(\ft{q}{\rho}),\qquad
\sigma=\sqrt{1-\ft{q^2}{\rho^2}}.
\ee
To be precise, $\sigma$ can be uniquely solved once we made the Ellis wormhole ansatz for the scalar $\phi$.
If we turn off  the charge, the remainder of the equations implies that $h=1$ giving rise to the standard Ellis wormhole. Turning on the charge, the two remaining independent equations are
\bea
&&Q^2Z+4\rho^2(h-1)+(4\rho^3-4q^2\rho)h'=0,\label{eq1}\\
&&-Q^2Z+(4\rho^3-2q^2\rho)h'+(2\rho^4-2q^2\rho^2)h''=0.\label{eq2}
\eea
Eliminating $Q$ from these two equations gives rise to a second-order differential equation of $h(\rho)$ which can be fully solved, giving
\be
h=1 + \fft{c_1}{\rho^2} + \fft{c_2 \sqrt{\rho^2-q^2}}{\rho^2}\,,
\ee
where the $(c_1,c_2)$ are two integration constants.  We can then substitute the solution back to \eqref{eq1}
and \eqref{eq2} and solve for $Z$ as a function of $\rho$.  It follows from \eqref{phi} that we have $\rho=q/\sin(\fft\phi 2)$ and hence
\be
Z(\phi)=\gamma_1 \cos\phi + \gamma_2 \sin\phi\,,\label{zphires}
\ee
where $\gamma_1=c_1/Q^2$ and $\gamma_2 = -4c_2/Q^2$.  Note that we redefined the constants $(c_1,c_2)$ to $(\gamma_1,\gamma_2)$ such that $Q$ does not appear in $Z(\phi)$ and hence it can be appropriately interpreted as an integration constant of the solution.  In other words, the solution for the $h$ now is expressed as
\be\label{whh}
h=1-\ft{\gamma_2 Q^2\sigma}{4q\rho}+\ft{\gamma_1 Q^2}{4 \rho^2}.
\ee
Finally, we need to verify the equation of motion of the phantom $\phi$ and it is satisfied.

To conclude, for the general class of the theory \eqref{Ztheory}, there exists one linear combination \eqref{sigma} of the equations that is independent of the charge for the spherically-symmetric and static ansatz.  This allows us to adopt the Ellis wormhole ansatz to solve \eqref{sigma}, we can then completely determine not only the $Z$, given in (\ref{zphires}), but also the solution
\bea\label{theo}
&&ds^2=-h dt^2+(\sigma^2 h)^{-1}d\rho^2+\rho^2d\Omega_{2}^2,\qquad A=\xi(\rho) dt,\nn\\
&&h=1-\ft{\gamma_2 Q^2\sigma}{4q\rho}+\ft{\gamma_1 Q^2}{4 \rho^2},\quad\sigma=\sqrt{1-\ft{q^2}{\rho^2}},\quad \phi=2\arcsin(\ft{q}{\rho}),\quad \xi'=\ft{QZ}{\sigma \rho^2}.
\eea
In this solution, the $(q,Q)$ are the two integration constants parameterizing the mass and the electric charge.  The parameters $(\gamma_1,\gamma_2)$ are the parameters of the theory, entering the coupling function $Z(\phi)$.

\section{Exact solutions of the photon sphere}

In fact, we can solve the \eqref{cubic} and obtain the three roots, which are given by
\bea\label{rr}
&&r_1=\big(-\ft{p_1}{2}+\sqrt{(\ft{p_1}{2})^2+(\ft{p_2}{3})^2}\big)^{\ft{1}{3}}+\big(-\ft{p_1}{2}-\sqrt{(\ft{p_1}{2})^2+(\ft{p_2}{3})^2}\big)^{\ft{1}{3}}+\ft{\gamma_2 Q^2}{8q},\nn\\
&&r_2=\omega\big(-\ft{p_1}{2}+\sqrt{(\ft{p_1}{2})^2+(\ft{p_2}{3})^2}\big)^{\ft{1}{3}}+\omega^2\big(-\ft{p_1}{2}-\sqrt{(\ft{p_1}{2})^2+(\ft{p_2}{3})^2}\big)^{\ft{1}{3}}+\ft{\gamma_2 Q^2}{8q},\nn\\
&&r_3=\omega^2\big(-\ft{p_1}{2}+\sqrt{(\ft{p_1}{2})^2+(\ft{p_2}{3})^2}\big)^{\ft{1}{3}}+\omega\big(-\ft{p_1}{2}-\sqrt{(\ft{p_1}{2})^2+(\ft{p_2}{3})^2}\big)^{\ft{1}{3}}+\ft{\gamma_2 Q^2}{8q},
\eea
with
\be\nn
\omega=\ft{-1+\sqrt{3}i}{2},\qquad p_1=q^2+\ft{\gamma_1 Q^2}{2}-\ft{3\gamma_2^2 Q^4}{64 q^2},\qquad p_2=\ft{\gamma_2 Q^2}{256 q^3}(64q^4+16q^2 Q^2-\gamma_2^2 Q^4).
\ee
Note that $r_1$ is a real root if all the parameters are real. Moreover, it can be confirmed that $r_1$ satiesfies the \eqref{con2} so it is a photon sphere. The requirement of the $r_2$ and $r_3$ to be real is quite complicated. We would like to show it pithily by defining some parameters
\bea
&&a=-9\gamma_2(4\gamma_1^2+3\gamma_2^2), b=32q^2\gamma_1(16\gamma_1^2+9\gamma_2^2), c=384q^4(\gamma_1^2+3\gamma_2^2),d=6144q^6\gamma_1,e=4096q^8\nn\\
&&\delta_1=c^2-3b d+12 a e,\quad \delta_2=2c^3-9b c d+27a d^2+27b^2 e-72 a c e,\quad m_1=\sqrt[3]{\delta_2+\sqrt{\delta_2^2-4\delta_1^3}},\nn\\
&&\delta_3=\sqrt[3]{2}\ft{\delta_1}{3a m_1}+\ft{m_1}{3\sqrt[3]{2}a},\quad m_2=-\ft{b^2}{4a^2}-\ft{2c}{3a},\quad m_3=\ft{1}{2}\sqrt{m_2+\delta_3}\quad m_4=\ft{4bc}{a^2}-\ft{b^3}{a^3}-\ft{8d}{a}, \nn\\
&&m_5=\ft{1}{2}\sqrt{2m_2-\delta_3+\ft{m_4}{8m_3}}, \qquad m_6=-\ft{b}{4a}.
\eea
Use those parameters, the requirement of $r_2$ and $r_3$ to be real is that
\be\label{con3}
Q\geq \sqrt{m_3+m_5+m_6},
\ee
where equivalence means $r_2=r_3$. Note that $r_2>r_3$ while they are unequal, it means that $r_2$ is a photon sphere and $r_3$ is a stable photon orbits. Comparing \eqref{bhcon} with
\be
\sqrt{m_3+m_5+m_6}<\ft{2\sqrt{2}}{\gamma_2}\sqrt{(\gamma_1 q^2+\sqrt{(\gamma_1^2+\gamma_2^2)q^4})}.
\ee
It shows that the black hole solution always has two photon spheres, and the wormhole solution has one or two photon spheres. The photon spheres locate at the outside of horizons, which can be proofed by showing that $h(r_1)$ and $h(r_2)$ are positive.

\end{document}